\documentclass{article}
\usepackage[utf8]{inputenc}
\usepackage{amsmath,amssymb}
\usepackage{amsthm}

\usepackage{centernot}
\usepackage[dvipsnames]{xcolor}
\usepackage[normalem]{ulem}
\usepackage{indentfirst}
\usepackage{booktabs}

\usepackage{array}
\renewcommand{\arraystretch}{1.5}

\usepackage{soul,xcolor}
\setstcolor{red}

\usepackage{graphicx}
\usepackage{float}
\usepackage{siunitx}
\usepackage{subcaption}

\usepackage[%
    colorlinks=false,
    pdfborder={0 0 0},
    linkcolor=black
]{hyperref}

\usepackage[stable]{footmisc}

\usepackage{geometry}
\usepackage[title]{appendix}

\usepackage[style=numeric-comp, maxcitenames=2, sorting=none]{biblatex}
\usepackage{bm}
\title{Formulations for scalar boundedness in simulations of turbulent compressible multi-component flows using high-order finite-difference methods}
\usepackage{authblk}

\newcommand\blfootnote[1]{%
  \begingroup
  \renewcommand\thefootnote{}\footnote{#1}%
  \addtocounter{footnote}{-1}%
  \endgroup
}
\author[1,*]{Ye Wang}
\author[2]{Armin Wehrfritz}
\author[1]{Evatt R. Hawkes}

\affil[1]{\small\textit{School of Mechanical and Manufacturing Engineering, The University of New South Wales, Sydney, Australia}}
\affil[2]{\small\textit{Department of Mechanical and Materials Engineering, University of Turku, Finland}}
\date{\vspace{-5ex}}

\renewenvironment{abstract}
 {\quote\small\noindent\rule{\linewidth}{.5pt}\par\smallskip
  {\raggedright\bfseries\abstractname\par}\medskip}
 {\par\noindent\rule{\linewidth}{.5pt}\endquote}

\providecommand{\keywords}[1]{{\textit{Keywords:}} #1}

\begin{document}

\maketitle

\begin{abstract}
    Preserving scalar boundedness is important for numerical schemes used in turbulent compressible multi-component flow simulations to prevent unphysical results and unstable simulations. However, ensuring scalar boundedness for high-order, low-dissipation numerical schemes poses challenges in highly under-resolved conditions due to inherent dispersion errors that generate spurious oscillations. Numerical dissipation is needed to mitigate these oscillations, but excessive dissipation negatively affects resolution. In this work, we propose formulations for high-order finite-difference schemes to preserve scalar boundedness without predefined bounds, while maintaining high accuracy and low numerical dissipation. The proposed formulations augment a non-dissipative numerical flux of a high-order central-difference scheme with an explicit dissipative numerical flux that adaptively switches between high-order and low-order formulations. Building on a deliberate choice of the non-dissipative flux, we construct two schemes using Jameson's artificial viscosity method and a monotonicity-preserving limiter as the dissipative flux. We examine the schemes in one-dimensional scalar advection problems and a three-dimensional temporal turbulent mixing-layer case involving sharp scalar gradients and under-resolved conditions, evaluating their accuracy, boundedness of species mass fractions, and numerical diffusivity. The scheme with the monotonicity-preserving limiter demonstrates superior performance. \blfootnote{$^*$Corresponding author (ye.wang7@unswalumni.com)}

\end{abstract}

\section{Introduction} \label{Sec_intro}
In direct numerical simulation and large-eddy simulation of turbulent compressible multi-component flows, it is recommended to use high-order, low-dissipation schemes to discretise the nonlinear convective terms~\cite{Ducros2000, Morinishi2010, Pirozzoli2010,Subbareddy2017}. However, in highly under-resolved conditions where sharp scalar gradients are not well resolved (such as thin mixing layers and flame fronts in reacting flows), dispersion errors of such schemes generate significant unphysical oscillations in scalar fields and may result in violations of the scalar bounds prescribed by initial and boundary values, i.e., violations of scalar boundedness~\cite{Herrmann2006, Matheou2016, Subbareddy2017, Ma2017}. 

Ensuring scalar boundedness is of great importance since scalar excursions (i.e., over-/undershoots) may lead to unphysical results and unstable simulations. For instance, oscillations induced by the numerical schemes may lead to the mass fraction of a chemical species, $Y$, exceeding its global physical bounds ($Y\in[0,1]$), which may even result in negative densities and/or temperature predictions exceeding the adiabatic flame temperature in combustion simulations~\cite{Subbareddy2017}. Simply clipping such over-/undershoots leads to significant conservation errors~\cite{Subbareddy2017,Mcdermott2015trimming}. To dampen these unphysical oscillations, numerical dissipation is often required, which may be introduced by upwind-schemes, filtering, or viscosity-like operators~\cite{Pirozzoli2011}. However, only first-order methods can ensure boundedness, which are highly dissipative and can lead to inaccurate predictions of turbulent mixing~\cite{Sharan2018}. Due to these conflicting requirements, there are very few schemes that guarantee scalar boundedness and simultaneously maintain an overall high-order accuracy, low artificial diffusion, and strong conservation properties, all of which are, however, essential in high-resolution turbulence simulations. Despite these challenges, several strategies to improve boundedness have been proposed for high-order numerical methods. These can be classified as flux corrected transport~\cite{Boris1973, Zalesak1979, Herrmann2006, Ma2017} or flux limiting schemes~\cite{Vanleer1973, Zhang2010, Verma2014, Subbareddy2017, Sharan2018}, and artificial viscosity methods~\cite{Jameson1981, Cook2007, Kawai2008, Kord2022}. 

The first class converts high-order numerical flux schemes to low-order numerical fluxes that preserve bounds wherever the prescribed bounds are violated (flux correction) or the flux limiter is activated (flux limiting). \textcite{Herrmann2006} constructed the bounded quadratic-upwind biased interpolation convective (BQUICK) scheme based on the flux correction method developed by \textcite{Boris1973, Zalesak1979}. It is a predictor-corrector algorithm that carries out \textit{a posteriori} boundedness checks on solutions advanced using the high-order QUICK scheme~\cite{Leonard1979QUICK} and then corrects out-of-bound values using a first-order upwind scheme. The BQUICK scheme succeeded in preserving scalar boundedness in turbulence simulations with minimal influence on the mixing field~\cite{Herrmann2006}, and has been applied to many applications including turbulent reacting simulations~\cite{Savard2015, Fillo2020}. However, the bounds need to be known as a prerequisite -- this can be a limitation because global bounds, while definable, may not always be meaningful or practical. For instance, in a domain with two unmixed streams, each having different minimum and maximum scalar values, a single global range may not capture local variations~\cite{Subbareddy2017}. \textcite{Ma2017} circumvented this issue by using the discrete minimum entropy principle~\cite{Tadmor1986} as a criterion to check for violations of local entropy bounds, which also implicitly ensures density and pressure positivities, and mass fractions within $[0,1]$. A common shortcoming of such flux correction strategies is each time stage is divided into two or more iterations, which would increase the computational cost considerably.

\textcite{Subbareddy2017} employed a linear-scaling limiter, based on the maximum-principle-satisfying approach developed by \textcite{Liu1996, Zhang2010}, for finite-volume methods in compressible flow simulations. The limiter selectively reverts high-order polynomial reconstructions to the first-order bounded reconstruction due to \textcite{Larrouturou1991} to preserve scalar boundedness. Although such linear-scaling limiters have been demonstrated to be effective in chemically reacting flows, see e.g., \textcite{Du2022}, also they require to prescribe the (global) scalar bounds. In lieu of such bound-preserving ideas, \textcite{Sharan2018} applied the monotonicity-preserving limiter devised by \textcite{Suresh1997} to a high-order centred finite-difference discretisation to enforce monotonicity of scalar solutions and thus boundedness. The monotonicity-preserving limiter checks the geometric monotonicity of high-order reconstructions~\cite{Suresh1997} and hence requires no inputs on the bounds, which indicates it can detect local scalar bounds that are not close to the global ones. Additionally, this limiter avoids degeneration of accuracy near local smooth extrema~\cite{Suresh1997}, which is a known drawback of total-variation-diminishing schemes~\cite{Osher1984}. However, unlike the decoupled reconstruction for passive scalars in incompressible flows~\cite{Sharan2018}, it is not straightforward to integrate the monotonicity-preserving limiter with a centred reconstruction for scalars in compressible flows, where, for example, the species mass fractions are nonlinearly coupled with the system. 

The second group augments high-order central-difference schemes with explicit artificial dissipation, using either dissipation operators or diffusive numerical fluxes that adaptively switch from high-order to low-order formulations at unbounded regions or near discontinuities to preserve boundedness. The idea of supplementing artificial viscosity coefficients with the physical ones originates from \textcite{VonNeumann1950} and was used, e.g., by \textcite{Cook2007, Kawai2008}. In their formulations, the boundedness of species mass fractions is ensured by a low-order artificial diffusion operator activated selectively via the Heaviside function, which is consistent with the physical mass fraction bounds, $[0,1]$. More recently, a similar method was devised by \textcite{Kord2023boundedness} for scalar boundedness and was applied in turbulent reacting simulations~\cite{Kord2022}. The same selective sensor, i.e., using global bounds as in Refs.~\cite{Herrmann2006, Cook2007}, was utilised, however, with the summation-by-parts operators due to \textcite{Mattsson2004} as the artificial dissipation. 

\textcite{Jameson1981} provided an alternative option for constructing explicit artificial dissipation using dissipative numerical fluxes, allowing for more straightforward implementation for schemes in a numerical flux form. In addition, the switch from high-order to low-order dissipative fluxes is typically controlled by sensors based on real-time flow solutions~\cite{Jameson1981, Oefelein1997, Ducros1999}, which has the prospect of localising the low-order dissipation in the vicinity of sharp jumps to remove oscillations and subsequently preserve boundedness without predefined bounds. \textcite{Jameson1981} originally proposed a first-order and a third-order dissipative flux, using a pressure-based shock sensor as the switching method for shock capturing. In a more recent study on high-speed reacting flows, \textcite{Sciacovelli2021} generalised the expressions for combinations of first-order and high-order dissipative fluxes (order $3$ to $9$), using shock sensors based on the classical Ducros sensor~\cite{Ducros1999} and Jameson's pressure-based sensor~\cite{Jameson1981}. Although Jameson's artificial viscosity method has been successful in shock capturing, its application to resolving sharp scalar gradients is relatively limited. For example, \textcite{Oefelein1997} combined first-order and up to fifth-order dissipative fluxes with both pressure-based and density-based switches, where the density switch proved useful in mixing-layer problems. Nevertheless, in shock-free turbulent flows with sharp scalar gradients, caution should be exercised to avoid being overly dissipative, as the above density-based sensor may capture regions near smooth, local extrema and turbulent fluctuations, leading to unwanted activation of the highly dissipative first-order flux and degeneration of accuracy. Although there are other types of sensors in the literature, such as the absolute jump detector used by \citeauthor{Pirozzoli2002} on density fields~\cite{Pirozzoli2002}, the minmod-based total-variation-bounded limiter due to \textcite{Cockburn1989}, or the smoothness indicator due to \textcite{Ren2003}, which was applied to identify material interfaces by~\textcite{Subbareddy2017}, the results for turbulent flows in particular are highly sensitive to the threshold values.

Therefore, even though various strategies to improve scalar boundedness have been proposed in the literature, a general approach that does not require predefined bounds and preserves high accuracy has not been reported, to the best of the authors' knowledge. In the present study, we aim to construct a formulation for high-order finite-difference methods that can preserve scalar boundedness without predefined bounds, while maintaining overall low numerical dissipation, high accuracy, and stability. Specifically, we consider the scalar transport equations for species mass fractions in compressible multi-component turbulent flows. The proposed formulations are based on the method of explicit artificial dissipation that combines a non-dissipative numerical flux with an explicit dissipative numerical flux. For the non-dissipative component, we use the flux developed in our previous study~\cite{Wang2025}, which has a physically consistent ‘skew-symmetric’ split form. Building on this, we then devise two dissipative fluxes: one follows Jameson's artificial viscosity method~\cite{Jameson1981}, using a density-based sensor, and the other is novel, integrating the monotonicity-preserving limiter due to \textcite{Suresh1997} to switch between high- and low-order dissipative fluxes. Neither scheme requires inputs on scalar bounds, allowing for adaptive boundedness preservation. We also evaluate the flux correction method with the proposed formulations. The properties and performance of the proposed schemes regarding boundedness, stability, accuracy, and numerical dissipation, are evaluated through numerical tests on one-dimensional scalar wave advection and three-dimensional turbulent mixing-layer problems involving sharp scalar gradients and under-resolved conditions. 

The remainder of the paper is organised as follows. Section~\ref{Sec_gover} summarises the governing equations. In Section~\ref{Sec_numtotal}, the numerical flux framework consisting of non-dissipative and explicit dissipative fluxes is introduced in Section~\ref{Sec-numericalflux-framework}, and then the bounded schemes are constructed in Section~\ref{Sec-scalarboundedness}, along with a summary of the schemes in Section~\ref{methodsummary}. Section~\ref{Sec_tests} presents the results from numerical tests, followed by concluding remarks in Section~\ref{Sec_con}.

\section{Governing equations} \label{Sec_gover}
We consider the three-dimensional Euler equations for compressible flows with $N_s$ species, as a well-defined model problem to study numerical methods for the convective terms in the Navier--Stokes equations. The conservative form of the Euler equations reads:
\begin{equation}
\begin{split}
  &\frac{\partial U}{\partial t}+\frac{\partial F_{j}}{\partial x_j}=0, \\[5pt]
  U=\left[\def\arraystretch{1.05}\begin{array}{c}\rho\\ \rho u \\ \rho v \\ \rho w \\ \rho E \\ \rho Y_\alpha\end{array}\right], \ \ \ 
  F_{j}=&\left[\def\arraystretch{1.05}\begin{array}{c}\rho u_j\\ \rho uu_j+p\delta_{1j} \\ \rho vu_j+p\delta_{2j} \\ \rho wu_j+p\delta_{3j} \\ \rho (k+e)u_j+pu_j \\ \rho Y_\alpha u_j\end{array}\right],\ \ \ 
  \alpha=1,\cdots,N_s-1,
  \label{EulerEqs}
\end{split}
\end{equation}
where $t$ denotes the time, $x_j$ are the spatial coordinates in a Cartesian coordinate system, $u_j$ is the velocity vector component in direction $j$ ($u=u_1, v=u_2, w=u_3$), $\delta_{ij}$ is the Kronecker delta, $\rho$ is the mixture density, $p$ is the mixture total pressure, $Y_\alpha$ is the mass fraction of species $\alpha$, and $N_s$ is the total number of species. Only $N_s-1$ species equations are solved and the mass fraction of species $N_s$ is determined by $\sum^{N_s}_{\alpha=1}Y_\alpha=1$, ensuring that the mass fractions of all species sum to unity by design. The specific total energy, $E=k+e$, is the sum of specific kinetic energy, $k=u_iu_i/2$ and mixture specific internal energy, $e$. We assume a calorically perfect gas with the ideal-gas equation of state:
\begin{equation}
    p=\rho\frac{R_u}{W}T=\rho(\gamma-1)e, \label{iglaw}
\end{equation}
where $T$ is the temperature and $R_u$ is the universal gas constant. The mean molar mass of the mixture, $W$, is defined as,
\begin{equation}
    W=\left( \sum_{\alpha=1}^{N_s}{\frac{Y_\alpha }{W_\alpha}} \right)^{-1}.  \label{molarmass}
\end{equation}
The mean specific heat ratio of the mixture, $\gamma$, is obtained from the relation:
\begin{equation}
    \frac{1}{\gamma-1}=\sum_{\alpha=1}^{N_s}{\frac{Y_\alpha}{\gamma_\alpha-1}}\frac{W}{W_\alpha}. \label{gammaandgammas}
\end{equation}
Here, $W_\alpha$ and $\gamma_\alpha$ respectively denote the molar mass and the specific heat ratio of species $\alpha$.

\section{Numerical methods} \label{Sec_numtotal}
In this section, we develop high-order finite-difference discretisations for the convective term in Eq.~\eqref{EulerEqs}, ${\partial F_{j}}/{\partial x_j}$. First, we propose a numerical flux in a framework that augments high-order non-dissipative schemes with explicit artificial diffusion in Section~\ref{Sec-numericalflux-framework}. Based on this framework, we then devise formulations for scalar boundedness in Section~\ref{Sec-scalarboundedness}. 

\subsection{Numerical flux formulation} \label{Sec-numericalflux-framework}
For compressible flows that may develop discontinuities, preserving global conservation in a discrete sense ensures that the discrete solution converges to a unique weak solution upon grid convergence, as stipulated by the Lax--Wendroff theorem~\cite{Lax1960}. In the numerical flux formulation, such as \textcite{Ducros2000}, the finite-difference operator is recast as the difference of numerical fluxes at adjacent intermediate nodes, which subsequently ensures global conservation through the telescopic property~\cite{Lax1960}. For first-order derivatives at grid point $m$, assuming one dimension and uniform grid spacing $\Delta x$ for simplicity, the numerical flux form is formulated as 
\begin{equation}
    \left.\frac{\partial F}{\partial x}\right\vert_{x=x_m}\approx\frac{\hat{F}_{m+\frac{1}{2}}-\hat{F}_{m-\frac{1}{2}}}{\Delta x}, \label{numericalflux}
\end{equation}
where $\hat{F}_{m\pm\frac{1}{2}}$ denotes the numerical flux at the interface/flux point $m\pm\frac{1}{2}$ between grid points $m$ and $m\pm1$.

High-order central-difference approximations are widely considered the gold standard for direct numerical simulations and highly resolved large-eddy simulations of compressible turbulence flows due to their low numerical dissipation and subsequently accurate solutions~\cite{Kennedy1994,Ducros2000, Morinishi2010, Pirozzoli2010}. However, in highly under-resolved simulations, numerical dissipation is essential to stabilise the simulation and inhibit unphysical oscillations~\cite{Ducros2000, Cook2004, Pirozzoli2011, Chandrashekar2015, Ranocha2017}. In the present study, we use the numerical flux formulation to selectively add artificial dissipation by formulating the flux in a non-dissipative part, $\hat{F}^{c}$, and a dissipative part, $\hat{F}^{d}$, given as 
\begin{equation}
    \hat{F}_{m+\frac{1}{2}}=\hat{F}^{c}_{m+\frac{1}{2}}-\alpha_{d,m+\frac{1}{2}}\hat{F}^{d}_{m+\frac{1}{2}}, \label{framework-multiplicative}
\end{equation}
where $\alpha_d \in [0,1]$ is a factor which allows localising the dissipative flux, e.g., near discontinuities, also referred to as low-dissipation switch~\cite{Subbareddy2009, Subbareddy2017}.
This formulation is a common way to introduce explicit artificial dissipation, such as in Jameson's artificial viscosity method~\cite{Jameson1981} or upwind schemes using either Roe's or Lax--Friedrichs flux functions~\cite{Roe1981, Leveque1992}. The formulations regarding the non-dissipative and dissipative fluxes we use in this study are detailed in the following. The bounded schemes are proposed in this framework of explicit diffusive flux in the subsequent subsection.

\subsubsection{Non-dissipative numerical flux}
The construction of non-dissipative numerical fluxes for the convective terms in the multi-component Euler equations, Eq.~\eqref{EulerEqs}, uses the notion of ‘skew-symmetric’ split forms from Refs.~\cite{Blaisdell1996, Feiereisen1981, Kennedy2008}. \textcite{Pirozzoli2010} generalised such split forms with arbitrary orders of accuracy, reformulating them in a numerical flux formulation, referred to as split numerical flux, given as 
\begin{align}
    \hat{F}^{c}_{m+\frac{1}{2}}=2\sum_{l=1}^{L}{a_{L,l}\sum_{k=0}^{l-1}{\widetilde{(f,g,h)}_{m-k,l}}}, \label{M0}
\end{align}
where $a_{L,l}$ are the coefficients in standard explicit central-difference approximations of the first-order derivative with a formal order of accuracy of $2L$. For numerical tests in this study, eighth-order central-difference schemes are used with coefficients: $a_{4,1}=4/5$, $a_{4,2}=-1/5$, $a_{4,3}=4/105$, $a_{4,4}=-1/280$. The operator $\widetilde{(f,g,h)}_{m-k,l}$ is a two-point discrete averaging operator, which varies for convective terms in different split forms~\cite{Pirozzoli2010,Coppola2019}. For the convective terms in the continuity, momentum, energy, and species equations, we adopt split numerical fluxes with their respective two-point averaging operators given by 
\begin{equation}
\begin{split}
    \frac{\partial \rho u}{\partial x}:\ \widetilde{(\rho,u,1)}_{m-k,l}=& \frac{\rho_{m-k}+\rho_{m-k+l}}{2}\frac{u_{m-k}+u_{m-k+l}}{2}, \\[3pt]
    \frac{\partial \rho u u}{\partial x}:\ \widetilde{(\rho,u,u)}_{m-k,l}=& \frac{\rho_{m-k}+\rho_{m-k+l}}{2}\frac{u_{m-k}+u_{m-k+l}}{2}\frac{u_{m-k}+u_{m-k+l}}{2}, \\[3pt]
    \frac{\partial p}{\partial x}:\ \widetilde{(p,1,1)}_{m-k,l}=& \frac{p_{m-k}+p_{m-k+l}}{2}, \\[3pt]
    \frac{\partial \rho k u}{\partial x}:\ \widetilde{(\rho,k,u)}_{m-k,l}=& \frac{\rho_{m-k}+\rho_{m-k+l}}{2}\frac{u_{m-k}u_{m-k+l}}{2}\frac{u_{m-k}+u_{m-k+l}}{2}, \\[3pt]
    \frac{\partial \rho e u}{\partial x}:\ \widetilde{(\rho,e,u)}_{m-k,l}=& \frac{(\rho e)_{m-k}+(\rho e)_{m-k+l}}{2}\frac{u_{m-k}+u_{m-k+l}}{2}, \\[3pt]
    \frac{\partial p u}{\partial x}:\ \widetilde{(p,u,1)}_{m-k,l}=& \frac{p_{m-k+l}u_{m-k}+p_{m-k}u_{m-k+l}}{2}, \\[3pt]
    \frac{\partial \rho Y_\alpha u}{\partial x}:\ \widetilde{(\rho,Y_\alpha,u)}_{m-k,l}=& \frac{(\rho Y_\alpha)_{m-k}+(\rho Y_\alpha)_{m-k+l}}{2}\frac{u_{m-k}+u_{m-k+l}}{2},\ \alpha=1,\ldots,N_s-1\ . \label{M0-details}
\end{split}
\end{equation}
These numerical fluxes in the continuity, momentum, and energy equations, constructed by \textcite{Kuya2018, Shima2021}, have shown the ability to maintain kinetic-energy-preserving and entropy-preserving properties, and to preserve pressure equilibrium for single-component flows at non-uniform density conditions~\cite{Shima2021}. Our recent work~\cite{Wang2025} further demonstrated improved stability with these split fluxes for multi-component flows, compared to most other schemes, such as those in Refs.~\cite{Kennedy2008,Pirozzoli2010,Kuya2018}. For the species equation, we showed in a previous study~\cite{Wang2025} that the quadratic-split numerical flux can preserve mass fraction uniformity and temperature equilibrium, resulting in more accurate solutions. Consequently, this group of numerical fluxes (Eq.~\eqref{M0-details}) enables good numerical stability with zero artificial dissipation and consistently maintains the aforementioned physical properties, i.e., being physically consistent. Therefore, they are selected as the non-dissipative component, $\hat{F}^{c}$, in the present study. For a detailed discussion of these formulations and their implications, interested readers are referred to our recent work~\cite{Wang2025}.

\subsubsection{Dissipative flux} \label{dissflux}
We then adopt the local Lax--Friedrichs flux (LLF), or Rusanov flux~\cite{Rusanov1962} formulation for the dissipative flux term, given as 
\begin{equation}
    \hat{F}^{d}_{m+\frac{1}{2}}=\frac{1}{2}\lambda_{\max,m+\frac{1}{2}}(U^{R}_{m+\frac{1}{2}}-U^{L}_{m+\frac{1}{2}}),\ \ \ \lambda_{max,m+\frac{1}{2}}=\max_{L,R}(|u|+c),
    \label{M1-general}
\end{equation}
where $U^R$ and $U^L$ are reconstructions of the conserved variables $U$ in Eq.~\eqref{EulerEqs} using upwind-biased operators on the same stencil as the non-dissipative term. The LLF-form dissipative flux has been applied in Refs.~\cite{Chandrashekar2015, Ranocha2017, Gassner2016} as a stabilisation term because this flux guarantees kinetic energy dissipation when applied with a kinetic-energy-preserving flux, such as the non-dissipative flux we adopt (Eqs.~\eqref{M0} and \eqref{M0-details}). By applying the LLF dissipative flux (Eq.~\eqref{M1-general}) on the same stencil as the non-dissipative flux (Eq.~\eqref{M0}), the complete scheme (Eq.~\eqref{framework-multiplicative}) leads to $(2L-1)^\text{th}$-order accuracy, as identified in Ref.~\cite{Sciacovelli2021}. Since we use the eighth-order ($L=4$) explicit central-difference approximation, upwind-biased reconstructions of the seventh-order dissipative flux are used, given as 
\begin{equation}
\begin{split}
    U^{7th,L}_{m+\frac{1}{2}}=\sum_{l=1}^{7}b_{l}U_{m-4+l}&,\ \ \ U^{7th,R}_{m+\frac{1}{2}}=\sum_{l=1}^{7}b_{l}U_{(m+1)+4-l}, \\[5pt]
    b_{1}=-\frac{3}{420},\ b_{2}=\frac{25}{420},\ b_{3}=-\frac{101}{420}&,\ b_{4}=\frac{319}{420},\ b_{5}=\frac{214}{420},\ b_{6}=-\frac{38}{420},\ b_{7}=\frac{4}{420}. \label{7th-order}
\end{split}
\end{equation}
This seventh-order dissipative flux acts similarly to the eighth-order dissipation operator in the form $\partial^8U/\partial x^8$ used by \textcite{Larsson2007}. The effective range of these high-order dissipation terms is highly skewed towards the largest wavenumbers to remove under-resolved modes with minimal impact on the largest scales~\cite{Larsson2007}. Due to this property, \textcite{Sciacovelli2021} demonstrated the potential of using such high-order dissipation terms as implicit sub-grid regularisation for large-eddy simulations.

Jameson's artificial viscosity method suggests that blending a low-order dissipative flux with a high-order dissipative flux such as Eq.~\eqref{M1-general} can reduce oscillations near shocks and contact discontinuities~\cite{Jameson1981, Ducros2000, Oefelein1997, Sciacovelli2021}. A similar approach blending high-order and low-order dissipation operators has also been used to preserve scalar boundedness~\cite{Cook2007, Kord2023boundedness}. In the following section, we will adopt this blending strategy to construct bounded schemes.

\subsection{Scalar boundedness} \label{Sec-scalarboundedness}
In compressible and/or multi-component flows, unphysical excursions (over- or undershoots) in scalar quantities such as temperature or species mass fractions have detrimental effects on the solution and stability of the simulation. Therefore, we next explore methods which can not only ensure scalar quantities stay within their global physical bounds (e.g., $Y_\alpha \in [0,1]$ or $\rho > 0$), but also, where possible, within bounds determined by local conditions, e.g., the minimum and maximum values around sharp scalar gradients. This is particularly important for species mass fractions, which we focus on in the following by constructing schemes for mass fraction boundedness. We also anticipate that the schemes can mitigate unphysical oscillations in density and temperature. We will build two bounded schemes upon the framework of explicit dissipative flux (Eq.~\eqref{framework-multiplicative}) and use the strategy of blending high-order and low-order dissipative fluxes based on Jameson's artificial viscosity method~\cite{Jameson1981}.

\subsubsection{Jameson's artificial viscosity method}
Jameson's artificial viscosity method~\cite{Jameson1981, Oefelein1997, Ducros2000, Sciacovelli2021} uses solution-adaptive blends of low- and high-order dissipative fluxes to eliminate oscillations near sharp scalar gradients or discontinuities, having the potential of preserving scalar boundedness without predefined bounds as an input. The low-order dissipation term, typically the first-order upwind reconstruction, is localised using highly selective sensors near sharp jumps to remove oscillations and preserve scalar boundedness, whereas the high-order dissipation term is applied in the rest of the domain to suppress high-wavenumber modes. 

Here, we build a formulation for scalar boundedness upon Jameson's artificial viscosity method, and evaluate it in numerical tests (cf.~Section~\ref{Sec_tests}) together with another bounded scheme proposed in Section~\ref{proposed}. We take the combination of first-order and seventh-order dissipation terms from \textcite{Sciacovelli2021}, where the latter results in the same form as Eq.~\eqref{7th-order}. Unlike Ref.~\cite{Sciacovelli2021}, we focus on scalar gradients in this paper instead of shocks. Hence, we adopt the density-based sensor~\cite{Oefelein1997,Hill2006} to switch between the low- and high-order dissipation terms, as this sensor effectively detects large density gradients often associated with large mass fraction gradients. The final expression is given as 
\begin{equation}
\begin{split}
    &\hat{F}^{d}_{m+\frac{1}{2}}=\frac{1}{2}\lambda_{\max,m+\frac{1}{2}}\bigl[\epsilon_{2,m+\frac{1}{2}}(U_{m+1}-U_{m})-\epsilon_{8,m+\frac{1}{2}}(U^{7th,R}_{m+\frac{1}{2}}-U^{7th,L}_{m+\frac{1}{2}})\bigr], \\[5pt]
    &\epsilon_{2,m+\frac{1}{2}}=k_2\max(\nu_m,\nu_{m+1}),\ \ \ \epsilon_{8,m+\frac{1}{2}}=\max(0,1-6\epsilon_{2,m+\frac{1}{2}}), \\[5pt]
    &\nu_m = \frac{|\rho_{m+1}-2\rho_{m}+\rho_{m-1}|}{|\rho_{m+1}+2\rho_{m}+\rho_{m-1}|},
    \label{M2-JS}
\end{split}
\end{equation}
where the first term and second term in the square brackets $\bigl[\cdots\bigr]$ are the first-order and seventh-order dissipative fluxes, respectively, $\nu$ is the density-based sensor, and we set the user-specified parameter $k_2=0.5$ for all computations in Section~\ref{Sec_tests}. Except for the sensor, the final dissipative flux is the same as the formulation by \textcite{Sciacovelli2021}.

Therefore, the complete scheme consists of the non-dissipative flux (Eq.~\eqref{M0}) and the dissipative flux (Eq.~\eqref{M2-JS}), and is hereafter referred to as M2-JS.

\subsubsection{Monotonicity-preserving flux} \label{proposed}
With state-of-the-art shock sensors Jameson's artificial viscosity method has been precise in targeting shocks and thus in localising the low-order dissipation. However, for shock-free turbulent flows with sharp scalar gradients, caution should be exercised as the above density-based sensor may lead to excessive first-order dissipation near smooth extrema and the subsequent accuracy degeneration. The accuracy of some other sensors, such as jump detectors~\cite{Pirozzoli2002} and smoothness indicators~\cite{Ren2003}, is highly dependent on the threshold values in turbulent simulations. 

To address these issues, we consider the monotonicity-preserving (MP) limiter~\cite{Suresh1997,Sharan2018} a promising approach. \textcite{Suresh1997} derived the MP limiter for high-order schemes with upwind-biased reconstructions to preserve the monotonicity of solutions without affecting accuracy at smooth extrema. This limiter sets up a constraint that stipulates upper and lower limits for reconstructions at the interfaces; satisfying this constraint ensures a monotone solution remains monotone with Runge--Kutta time stepping (under a CFL restriction). The MP limiter is given as 
\begin{equation}
    w_{m+\frac{1}{2}}=\mathrm{median}(w^{\mathrm{high}}_{m+\frac{1}{2}},w^{\max}_{m+\frac{1}{2}},w^{\min}_{m+\frac{1}{2}}),
\end{equation}
where $w^{\mathrm{high}}_{m+\frac{1}{2}}$ represents the original high-order reconstruction of solution $w$, $w^{\max}_{m+\frac{1}{2}}$ and $w^{\min}_{m+\frac{1}{2}}$ respectively denote the upper and lower limits, $w_{m+\frac{1}{2}}$ is the limited reconstruction, and the median of the three terms gives the term that lies between the other two. The upper and lower limits are determined from low-order reconstructions that preserve monotonicity, and the interval between these limits is larger near smooth extrema to avoid loss of accuracy. \textcite{Suresh1997} defined curvature measures based on geometric analysis to distinguish between monotone data and extrema, and designed the enlarged limits. By applying the MP limiter, the original high-order reconstructions that lie within the lower or upper limits remain unaltered, whereas those lying outside are constrained to these limits. Accordingly, the limited reconstruction ensures monotonicity of the solution at the next time stage.

This MP limiter shows potential as both a low-order flux and a solution-adaptive sensor in Jameson's artificial viscosity method, for the following reasons. First, its lower and upper limits are essentially low-order reconstructions that ensure monotonicity and thus boundedness; hence, they can work as the low-order dissipation term to eliminate oscillations and preserve boundedness. Second, the median function, which adaptively switches between low- and high-order reconstructions, acts in a manner similar to the sensors or switching methods. Furthermore, the MP limiter might outperform Eq.~\eqref{M2-JS} as the MP limiter avoids unnecessary activation of low-order reconstructions at smooth extrema and provides less dissipative low-order reconstructions than the first-order upwind reconstructions, as elaborated by \textcite{Suresh1997}. However, it remains to be determined which switching method more precisely localises low-order dissipation near sharp scalar gradients. 

In light of these considerations, we apply the monotonicity-preserving limiter to the explicit diffusive flux framework and devise a new bounded scheme. The dissipative part is thus formulated as 
\begin{equation}
\begin{split}
    &\hat{F}^{d}_{m+\frac{1}{2}}=\frac{1}{2}\lambda_{\max,m+\frac{1}{2}}(U^{MP,R}_{m+\frac{1}{2}}-U^{MP,L}_{m+\frac{1}{2}}), \label{M2-MP}
\end{split}
\end{equation}
where $U^{MP,R}$ and $U^{MP,L}$ are the monotonicity-preserving reconstructions of the conserved variable $U$ using \citeauthor{Suresh1997}'s monotonicity-preserving limiter, expressed as
\begin{equation}
\begin{split}
    &U^{MP,L}_{m+\frac{1}{2}}=\mathrm{median}(U^{\mathrm{high},L}_{m+\frac{1}{2}},U^{\min,L}_{m+\frac{1}{2}},U^{\max,L}_{m+\frac{1}{2}}), \\[6pt]
    &U^{MP,R}_{m+\frac{1}{2}}=\mathrm{median}(U^{\mathrm{high},R}_{m+\frac{1}{2}},U^{\min,R}_{m+\frac{1}{2}},U^{\max,R}_{m+\frac{1}{2}}), \\[0.5pt]
\end{split}
\end{equation}
where $U^{\mathrm{high},L}$ and $U^{\mathrm{high},R}$ denote the high-order reconstructions (using the seventh-order scheme in Eq.~\eqref{7th-order} for the numerical tests in Section~\ref{Sec_tests}). The upper and lower limits for the left-state reconstruction, $U^{\max,L}$ and $U^{\min,L}$, are obtained following the procedure in \citeauthor{Suresh1997}'s monotonicity-preserving limiter as
\begin{equation}
\begin{split}
    &U^{\max,L}_{m+\frac{1}{2}} = \min\bigl[\max(U_{m},U_{m+1},U^{MD}_{m+\frac{1}{2}}),\ \max(U_{m},U^{UL,L}_{m+\frac{1}{2}},U^{LC,L}_{m+\frac{1}{2}})\bigr], \\[5pt]
    &U^{\min,L}_{m+\frac{1}{2}} = \max\bigl[\min(U_{m},U_{m+1},U^{MD}_{m+\frac{1}{2}}),\ \min(U_{m},U^{UL,L}_{m+\frac{1}{2}},U^{LC,L}_{m+\frac{1}{2}})\bigr], 
\end{split}
\end{equation}
where the quantities $U^{MD}$, $U^{UL,L}$, and $U^{LC,L}$ (MD, UL, and LC stand for median, upper limit, and large curvature, respectively) are given by 
\begin{align}
    &U^{MD}_{m+\frac{1}{2}} = \frac{1}{2}(U_{m}+U_{m+1})-\frac{1}{2}d^{M4}_{m+\frac{1}{2}}, \label{U-MD} \\[5pt]
    &U^{UL,L}_{m+\frac{1}{2}} = U_{m}+\alpha_{mp}(U_{m}-U_{m-1}), \\
    &U^{LC,L}_{m+\frac{1}{2}} = U_{m}+\frac{1}{2}(U_{m}-U_{m-1})+\frac{4}{3}d^{M4}_{m-\frac{1}{2}},
\end{align}
with
\begin{align}
    &d^{M4}_{m+\frac{1}{2}} = \mathrm{minmod}(4d_{m}-d_{m+1},4d_{m+1}-d_{m},d_{m},d_{m+1}), \label{dM4} \\[2pt]
    &d_{m} = U_{m+1}-2U_{m}+U_{m-1}. \label{dM4-d}
\end{align}
The right-state formulations are symmetric counterparts of the left-state formulations about the interface $m+\frac{1}{2}$, and are provided as follows:
\begin{equation}
\begin{split}
    &U^{\max,R}_{m+\frac{1}{2}} = \min\bigl[\max(U_{m},U_{m+1},U^{MD}_{m+\frac{1}{2}}),\ \max(U_{m+1},U^{UL,R}_{m+\frac{1}{2}},U^{LC,R}_{m+\frac{1}{2}})\bigr], \\[5pt]
    &U^{\min,R}_{m+\frac{1}{2}} = \max\bigl[\min(U_{m},U_{m+1},U^{MD}_{m+\frac{1}{2}}),\ \min(U_{m+1},U^{UL,R}_{m+\frac{1}{2}},U^{LC,R}_{m+\frac{1}{2}})\bigr], \\[2pt]
    &U^{UL,R}_{m+\frac{1}{2}} = U_{m+1}+\alpha_{mp}(U_{m+1}-U_{m+2}), \\
    &U^{LC,R}_{m+\frac{1}{2}} = U_{m+1}+\frac{1}{2}(U_{m+1}-U_{m+2})+\frac{4}{3}d^{M4}_{m+\frac{3}{2}}.
\end{split}
\end{equation}
We use the parameter $\alpha_{mp}=2$ for all computations in Section~\ref{Sec_tests}. In the original monotonicity-preserving scheme \cite{Suresh1997}, $\alpha_{mp}$ determines a CFL constraint for explicit Runge--Kutta time methods, $\mathrm{CFL}\leq1/(1+\alpha_{mp})=0.33$, and thus we keep the constraint satisfied in our numerical tests.

Hence, the complete scheme comprises the non-dissipative flux (Eq.~\eqref{M0}) and the dissipative flux (Eq.~\eqref{M2-MP}), and is hereafter referred to as M2-MP. The novelty of M2-MP lies in its formulation of a new solution-adaptive, bounded, dissipative flux based on the MP scheme and the integration with our previously proposed split non-dissipative flux.

\subsubsection{\textit{A posteriori} flux correction}
The BQUICK scheme due to \textcite{Herrmann2006} and entropy-stable flux correction due to \textcite{Ma2017} have been shown to ensure scalar boundedness with minimal modifications to the flow fields. The approach is based on an \textit{a posteriori} check on the solution at the next time stage, which was advanced by any numerical flux schemes. The check is based on a suitable criterion to detect violations of scalar boundedness, e.g., the discrete minimum entropy principle~\cite{Tadmor1986, Ma2017} or prescribed global bounds~\cite{Herrmann2006}. At grid points where the original numerical flux leads to a boundedness violation, the numerical flux at adjacent interfaces will be replaced by a low-order flux given as
\begin{align}
    \hat{F}^{\mathrm{low}}_{m+\frac{1}{2}}=\frac{1}{2}[F(U_m)+F(U_{m+1})]-\frac{1}{2}\lambda_{\max,m+\frac{1}{2}}(U_{m+1}-U_{m})\ , \label{1st-llf}
\end{align}
which is the first-order LLF flux formulation and also called the entropy-stable flux~\cite{Tadmor1984,Ma2017}. Finally, with the corrected numerical flux, the solution is re-advanced to the next time stage. Multiple iterations could be required, as at a neighbouring grid point, not all of its adjacent interfaces are modified by the low-order flux~\cite{Ma2017, Subbareddy2017}. 

The \textit{a posteriori} flux correction is suitable for use as a fail-safe option, as it offers flexible activation and ensures strict scalar boundedness. Therefore, we implement this correction step for the M2-MP scheme and evaluate its performance in numerical tests. The implementation is summarised in Table~\ref{tab:posteriorimplementation}, which follows Algorithm 2 due to \textcite{Ma2017}, with two modifications: 1) to minimise the impact only the numerical fluxes for the mass fraction transport equations are modified; 2) instead of using the discrete minimum entropy principle, we adopt the global bounds for mass fraction boundedness as the detection criterion as per Ref.~\cite{Herrmann2006} for efficiency. Hereafter, we refer to the scheme that augments the M2-MP scheme with this correction step as M3-MP.

\begin{table}[ht]
    \newcolumntype{L}[1]{>{\raggedright\arraybackslash}p{#1}}
    \setlength\tabcolsep{1pt}
    \small
    \renewcommand{\arraystretch}{1.2}
    \begin{tabular}{L{1.6cm}L{12.95cm}}
         \hline
          \multicolumn{2}{l}{\hspace*{-1mm} \textbf{for} each stage $k$ in Runge--Kutta time integration \textbf{do}} \\
          \hspace*{+5mm} step 1: & advance solution to the next stage $U^{k+1}$ using the M2-MP scheme; \\
          \hspace*{+5mm} step 2: & mark the unboundedness points if $Y^{k+1}_\alpha>Y^{\max}_\alpha$ or $Y^{k+1}_\alpha<Y^{\min}_\alpha$; \\
          \hspace*{+5mm} step 3: & at adjacent interfaces for the marked points, compute the difference between the low-order flux (Eq.~\eqref{1st-llf}) and the original high-order flux (M2-MP) for the mass fraction equations; \\
          \hspace*{+5mm} step 4: & for the marked points and their neighbours, correct the solution by the difference between the low-order flux and the original flux; \\
          \hspace*{+5mm} step 5: & go to step 2 for another iteration (optional); \\
          \multicolumn{2}{l}{\hspace*{-1mm} \textbf{end}} \\
         \hline
    \end{tabular}
    \caption{Algorithm of the \textit{a posteriori} flux correction step.}
    \label{tab:posteriorimplementation}
\end{table}

\subsection{Summary of schemes} \label{methodsummary}

In the explicit dissipative flux framework (Eq.~\eqref{framework-multiplicative}), we developed two bounded schemes, M2-JS and M2-MP. Both integrate solution-adaptive, bounded, dissipative fluxes with the split non-dissipative flux we previously formulated~\cite{Wang2025}. M2-JS and M2-MP differ in their dissipative flux formulations. The dissipative flux of M2-JS follows Jameson's artificial viscosity method~\cite{Jameson1981}, using a density-based sensor to switch the dissipative flux from high- to first-order upwind reconstructions. The dissipative flux of M2-MP employs the monotonicity-preserving limiter due to \textcite{Suresh1997} to construct a solution-adaptive, bounded, dissipative flux. 

In the following section, we evaluate the performance of these bounded schemes in numerical tests. To explain the role of each component in these schemes, we compare the bounded schemes with the unbounded, high-order dissipative scheme (M1) and the non-dissipative scheme (M0). The effect of the \textit{a posteriori} flux correction on the M2-MP scheme is also examined. All these schemes are summarised in Table~\ref{tab:testingmatrix}. Note that the factor $\alpha_d$ is taken either $0$ or $1$ for M0 or the other schemes. This study does not include cases with a detection switch for $\alpha_d$. We have tested Pirozzoli's jump detector~\cite{Pirozzoli2002}, where the results showed that it compromises the good boundedness property of the bounded schemes. Developing or identifying an appropriate switch for $\alpha_d$ is challenging and is beyond the scope of the present paper.

\begin{table}[ht]
    \centering
    \setlength\tabcolsep{5pt}
    \renewcommand{\arraystretch}{1.3}
    \begin{tabular}{l l c c c c}
         \hline 
         Scheme   & Description                                 & $\hat{F}^{c}$ & $\alpha_d$ & $\hat{F}^{d}$      & FC  \\
         \hline
         M0       & unbounded, non-dissipative scheme        & \eqref{M0}    & 0          & N/A                & off \\
         M1       & unbounded, high-order dissipative scheme & \eqref{M0}    & 1          & \eqref{M1-general} & off \\
         M2-JS    & bounded scheme using Jameson's artificial viscosity method & \eqref{M0}    & 1          & \eqref{M2-JS}      & off \\
         M2-MP    & bounded scheme using the monotonicity-preserving scheme & \eqref{M0}    & 1          & \eqref{M2-MP}      & off \\ 
         M3-MP    & M2-MP with \textit{a posteriori} flux correction  & \eqref{M0}    & 1          & \eqref{M2-MP}      & on  \\ 
         \hline
    \end{tabular}
    \caption{Numerical schemes to be tested. The components are based on the generic numerical flux framework (Eq.~\eqref{framework-multiplicative}). The number in the parentheses refers to the equation for the formulation, and ‘FC’ indicates the \textit{a posteriori} flux correction step.}
    \label{tab:testingmatrix}
\end{table}

\section{Numerical tests} \label{Sec_tests}
\subsection{One-dimensional advection} \label{1D_results}
In this section, we evaluate the performance of the proposed schemes on one-dimensional advection problems of scalar waves with uniform pressure and velocity. Similar to the study by \textcite{Ma2017}, we consider two cases, one with smooth initial conditions and another one with discontinuous ones. We consider a mixture consisting of four species –– $\mathrm{H_2}$, $\mathrm{H_2O}$, $\mathrm{O_2}$, and $\mathrm{N_2}$. The initial velocity, pressure, and density profiles for both cases are given as 
\begin{equation}
\begin{split}
    &u = 1,\ p = 1,\ \\
    & \rho = \rho_{\mathrm{H_2}}+\rho_{\mathrm{H_2O}}+\rho_{\mathrm{O_2}}+\rho_{\mathrm{N_2}},
    \nonumber
\end{split}
\end{equation}
where $\rho_{\alpha}=\rho_{\mathrm{sf},\alpha} Y_{\alpha}$ are the initial partial densities of species $\alpha$. The scaling factor, $\rho_{\mathrm{sf},\alpha}$, and species specific heat ratios, $\gamma_{\alpha}$, are set as 
\begin{equation}
\begin{split}
    (\rho_{\mathrm{sf},\mathrm{H_2}}, \rho_{\mathrm{sf},\mathrm{H_2O}}, \rho_{\mathrm{sf},\mathrm{O_2}}, \rho_{\mathrm{sf},\mathrm{N_2}}) &= (0.1, 0.8, 1.2, 1.0), \\
    (\gamma_{\mathrm{H_2}}, \gamma_{\mathrm{H_2O}}, \gamma_{\mathrm{O_2}}, \gamma_{\mathrm{N_2}}) &= (1.4, 1.33, 1.4, 1.4).
    \nonumber
\end{split}
\end{equation}
The species mass fractions, $Y_{\alpha}$ for $\alpha=\mathrm{H_2},\mathrm{H_2O},\mathrm{O_2}$, are specified as sinusoidal profiles for the case with smooth initial conditions, whereas they are initialised with sharp jumps for the case with discontinuous initial conditions, given as 
\begin{equation}
\begin{split}
    \mathrm{Smooth}&:Y_{\alpha} = 0.5(Y_{\mathrm{max},\alpha}+Y_{\mathrm{min},\alpha})+0.5(Y_{\mathrm{max},\alpha}-Y_{\mathrm{min},\alpha})\mathrm{sin}(2\pi x-\pi), \\
   \mathrm{Sharp}&:Y_{\alpha} =
    \begin{cases}
      Y_{\mathrm{max},\alpha} & 0.25<x<0.75 \\
      Y_{\mathrm{min},\alpha} & \text{otherwise},
    \end{cases}
    \nonumber
\end{split}
\end{equation}
where the maximum and minimum values are 
\begin{equation}
\begin{split}
    (Y_{\mathrm{max},\mathrm{H_2}},Y_{\mathrm{max},\mathrm{H_2O}},Y_{\mathrm{max},\mathrm{O_2}}) &= (0.8,0.0,0.17), \\
    (Y_{\mathrm{min},\mathrm{H_2}},Y_{\mathrm{min},\mathrm{H_2O}},Y_{\mathrm{min},\mathrm{O_2}}) &= (0.0,0.5,0.17),
    \nonumber
\end{split}
\end{equation}
and the computational domain is $x \in [0,1]$. Note that the $\mathrm{O_2}$ mass fraction is uniform in space for both cases, which allows assessing whether the numerical scheme preserves the uniformity in time. The last species, i.e., $\mathrm{N_2}$, is obtained from $Y_{N_s} = 1 - \sum_{\alpha = 1}^{N_s-1} Y_\alpha$. The mixture states resulting from the above setups have a maximum density ratio of approximately $3$, and a maximum temperature ratio of approximately $3.5$ for the smooth case and approximately $3$ for the sharp case. Hence, the sharp case has contact discontinuities, where the main numerical challenge for a high-order scheme is to preserve the scalar bounds, particularly for the species mass fractions, with low numerical dissipation. 

The numerical schemes listed in Table~\ref{tab:testingmatrix} are tested, including M0, M1, M2-JS, and M2-MP. The \textit{a posteriori} flux correction (M3-MP) is not employed in the 1D simulations, as the results are shown to be bounded well by M2-MP. The computations are carried out on a uniform grid with $\Delta x=0.01$, leading to $101$ grid points, and periodic boundary conditions. Time integration uses the three-stage third-order total-variation-diminishing (TVD) Runge--Kutta scheme due to \textcite{Shu1988}. M2-MP is shown to be stable and to exhibit good boundedness under the CFL constraint (see~Section~\ref{proposed}). The following results are reported at a Courant number of $0.01$, which is sufficiently small to ensure temporal convergence and to minimise time-integration errors when assessing spatial errors.

\subsubsection{Overall description}
\begin{figure}[!htb]
    \centering
    \includegraphics[keepaspectratio=true]{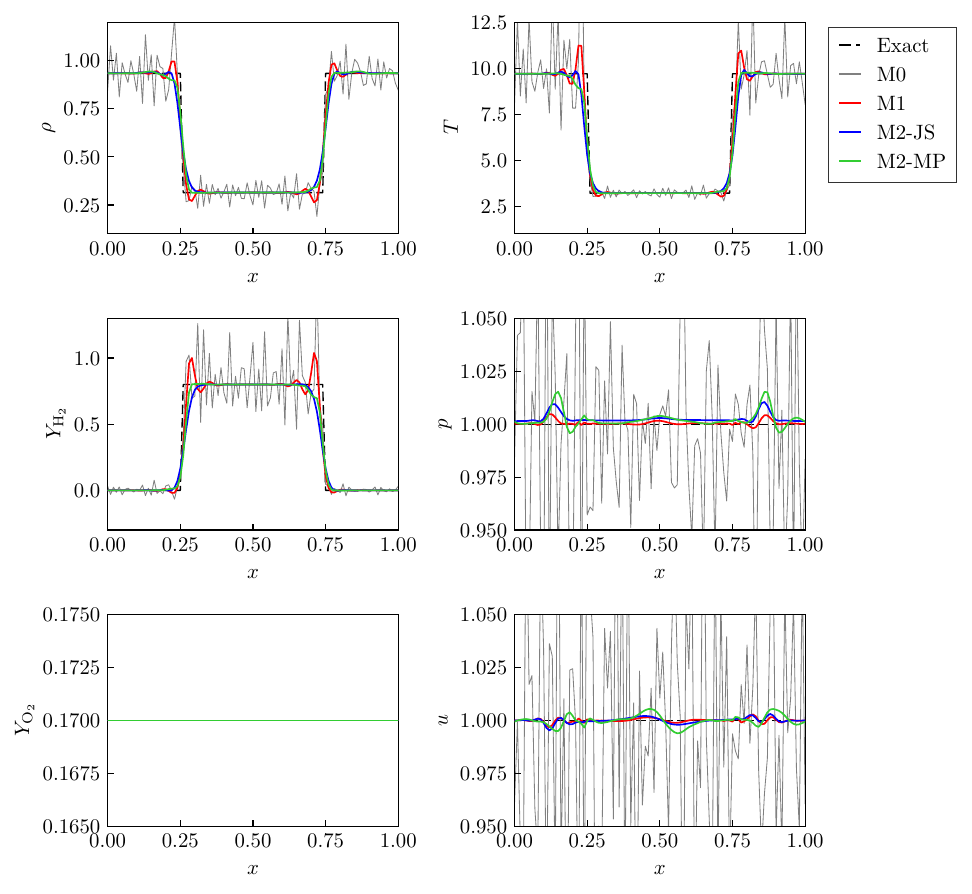}
    \vspace*{-7mm}
    \caption{Results of density, temperature, $\mathrm{H_2}$ mass fraction, $\mathrm{O_2}$ mass fraction, and velocity over one period for the one-dimensional advection case with sharp initial conditions.}
    \label{fig:adsharp} 
\end{figure}
Figure~\ref{fig:adsharp} shows the results computed by different numerical schemes for density, temperature, $\mathrm{H_2}$ mass fraction, pressure, $\mathrm{O_2}$ mass fraction, and velocity after one period of advection (i.e., $t=1$) for the case with sharp jump initial conditions. The exact solution is the initial condition. The M0 scheme generates high numerical oscillations in all variables except $Y_\mathrm{O_2}$. These oscillations are incurred by large dispersion errors of this high-order non-dissipative scheme. It also fails to maintain the equilibrium of pressure and velocity. The simulation becomes unstable soon after the first period and eventually blows up at time $t=1.44$. With the high-order dissipative flux term, the M1 scheme significantly cuts down the oscillations and therefore stabilises the simulation. Nevertheless, there still exist significant excursions in temperature and $\mathrm{H_2}$ mass fraction around the jumps. However, by adding locally low-order dissipative flux terms, the M2-JS and M2-MP schemes produce nearly bounded results in species mass fractions and temperature and generate only slight wiggles in the density distribution. It is also shown that the M2-MP scheme resolves a sharper interface than the M2-JS scheme, indicating that the M2-MP scheme adds overall less numerical dissipation. This is further analysed in Section~\ref{1D-boundedness}. The uniform $\mathrm{O_2}$ mass fraction profile is preserved precisely with errors on the order of $10^{-14}$. None of the schemes is able to strictly preserve the equilibrium of pressure and velocity. This is a known limitation of fully conservative formulations resolving discontinuities with jumps in specific heats~\cite{Abgrall2001, Johnsen2012}. However, for M2-JS and M2-MP the peak pressure oscillations remain below $3\%$ (considered small~\cite{Subbareddy2017}) and do not grow over time. Since the focus of the present paper is scalar boundedness, we do not further pursue strict pressure/velocity equilibrium here.

\begin{figure}[!htb]
    \centering
    \includegraphics[keepaspectratio=true]{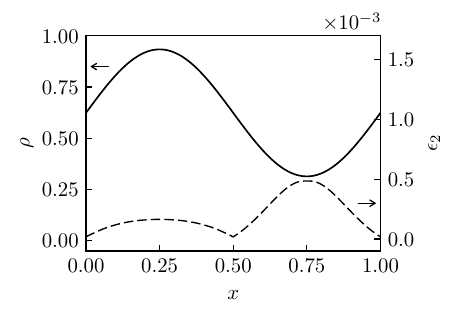}
    \vspace*{-3mm}
    \caption{Activation of the low-order term (dashed line) of the M2-JS scheme based on the density profile (solid line) after one-period advection for the case with smooth initial conditions.}
    \label{fig:detector} 
\end{figure}

For the case with smooth initial conditions, all numerical schemes yield results that are visually indistinguishable from the exact solution and are therefore not shown here. Quantitative errors are provided in Section~\ref{1D-convergence}. It is worth noting that the low-order dissipation term of the M2-MP scheme is not activated for this case, resulting in identical solutions to the M1 scheme. For the M2-JS schemes, however, the density sensor activates the low-order dissipation. As illustrated in Figure~\ref{fig:detector}, the activation region of the low-order term, represented by $\epsilon_2$, is nonzero and increases at local peaks in the density profile. This introduces unnecessary numerical dissipation to smooth solutions. Although the dissipation is small, it affects the accuracy and reduces the convergence rate, which is shown later in Section~\ref{1D-convergence}.

\subsubsection{Unboundedness error and interface thickness} \label{1D-boundedness}
\begin{figure}[!htb]
    \centering
    \includegraphics[keepaspectratio=true]{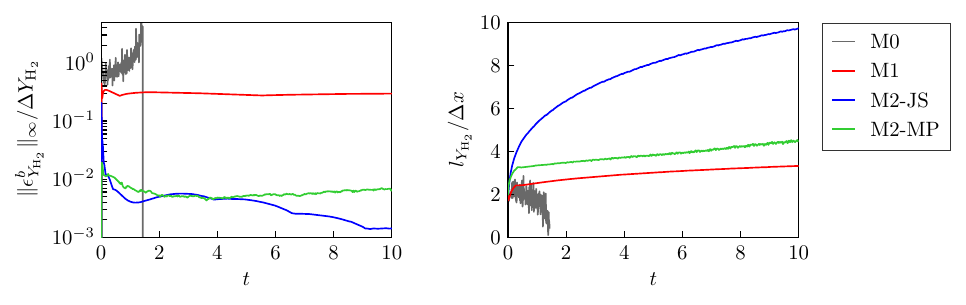}
    \vspace*{-7mm}
    \caption{Temporal evolution of normalised $\mathrm{H_2}$ mass fraction boundedness error and normalised numerical interface thickness for the case with sharp initial conditions.}
    \label{fig:ubdandthick}
\end{figure}

To study the effect of dissipative fluxes, we quantify the degree of local unboundedness and the degree of numerical diffusion for the case with sharp initial conditions. To assess the local unboundedness, we apply the method proposed by \textcite{Kord2023boundedness} to the $\mathrm{H_2}$ mass fraction, with lower and upper bounds of $0.0$ and $0.8$, respectively, resulting from the initial conditions. The unboundedness metric is thus defined as
\begin{equation}
\begin{split}
   \epsilon^b_{Y_\mathrm{H_2}} =
    \begin{cases}
      0 & \mathrm{if}\ 0.0\leq Y_\mathrm{H_2}\leq 0.8\\
      Y_\mathrm{H_2}-0.8 & \mathrm{if}\ Y_\mathrm{H_2}>0.8 \\
      0-Y_\mathrm{H_2} & \mathrm{if}\ Y_\mathrm{H_2}<0.0,
    \end{cases}
    \label{unboundedneserror}
    \nonumber
\end{split}
\end{equation}
from which we obtain the maximum error, i.e., the $\infty$-norm,
\begin{equation}
    \Vert\epsilon^b_{Y_\mathrm{H_2}}\Vert_\infty=\max\limits_{x=0,\cdots,1}{\epsilon^b_{Y_\mathrm{H_2}}}.
    \nonumber
\end{equation}
Temporal evolution of the maximum unboundedness error (normalised by the jump magnitude across the interface, $\Delta Y_\mathrm{H_2}=0.8$) is presented in the left plot in Figure~\ref{fig:ubdandthick}, where the vertical axis is plotted on a logarithmic scale. The degree of numerical diffusion is assessed by the normalised numerical interface thickness as defined in Ref.~\cite{Kawai2015}: 
\begin{equation}
   \dfrac{l_{Y_\mathrm{H_2}}}{\Delta x} = \dfrac{\frac{\Delta Y_\mathrm{H_2}}{\Delta x}}{\frac{\partial Y_\mathrm{H_2}}{\partial x}\vert_{\max}},
    \nonumber
\end{equation}
where the derivative is computed using the standard second-order explicit central difference approximation. The larger the value, the wider the interface and thus the higher the numerical diffusion. Temporal evolution of the normalised numerical interface thickness is shown in the right plot in Figure~\ref{fig:ubdandthick}. Comparing the schemes in Figure~\ref{fig:ubdandthick}, we observe a trade-off between the unboundedness error and numerical interface thickness. The M1 scheme maintains a large unboundedness error steadily in time while preserving a sharp interface, given that the thickness is only slightly wider compared to it at the initial time and also increases mildly with time. With the low-order dissipative fluxes in the M2-JS and M2-MP schemes, the unboundedness error is damped by two orders and the interface is thickened. With the M2-JS scheme, the interface thickness grows significantly faster over time, indicating increasingly more artificial diffusion is introduced. This also explains why its unboundedness error decreases with time. For the M2-MP scheme, the sharpness of the contact discontinuity is less smeared out, and the thickness grows at a slow, linear rate. The unboundedness error from the M2-MP scheme remains at a nearly constant value below $10^{-2}$. Comparing the two schemes, we find that the M2-MP scheme achieves a better balance between the two metrics.

\subsubsection{Convergence tests} \label{1D-convergence}
\begin{figure}[!t]
    \centering
    \includegraphics[keepaspectratio=true]{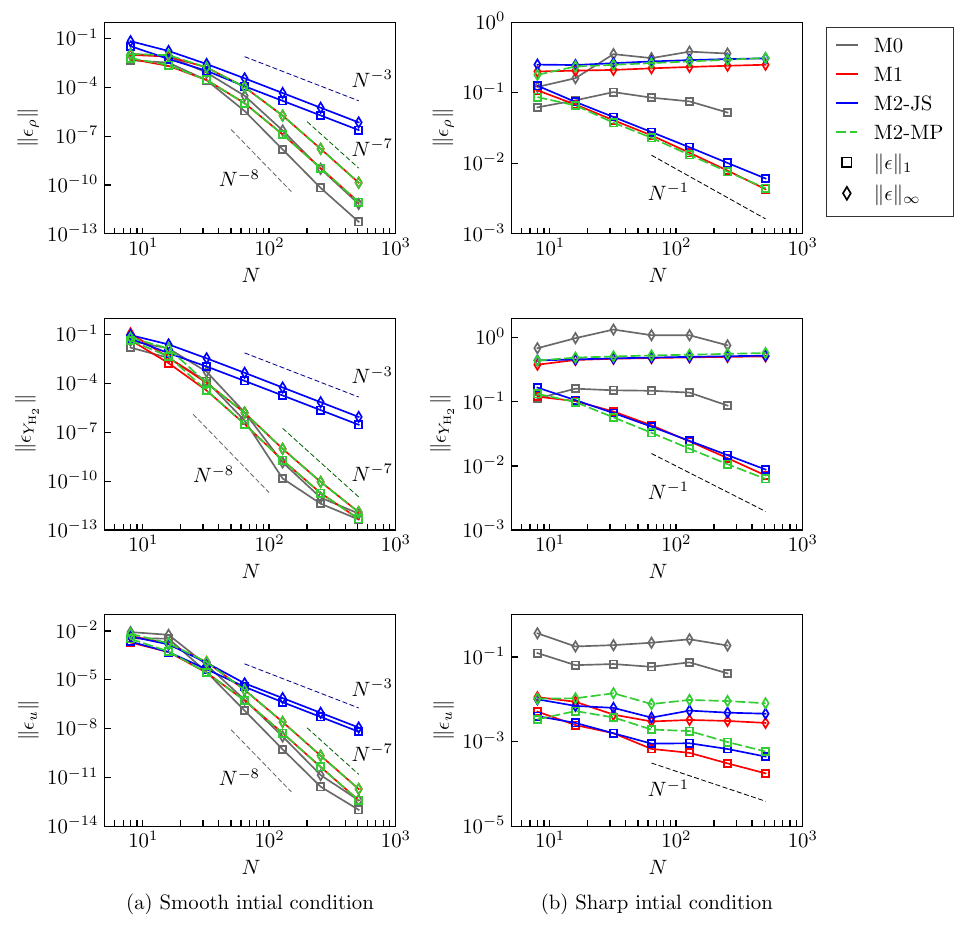}
    \vspace*{-7mm}
    \caption{The $L^1$ norm (labelled using $\square$) and $L^\infty$ norm (labelled using $\lozenge$) of errors in density, $\mathrm{H_2}$ mass fraction and velocity in computations for the cases with (a) smooth initial condition and (b) sharp initial condition.}
    \label{fig:convergence} 
\end{figure}
To verify the order of accuracy of the proposed schemes, a convergence study is carried out and results are shown in Figure~\ref{fig:convergence}, plotted on a logarithmic scale. Results (at $t=1$) in terms of the $L^1$ and $L^\infty$ norms of errors in density, $\mathrm{H_2}$ mass fraction, and velocity are shown. For the smooth solutions, as designed, the M0 and M1 schemes retain eighth-order accuracy and seventh-order accuracy, respectively. For the M2-JS scheme, nevertheless, due to the activation of low-order dissipation term on smooth solutions (cf.~Figure~\ref{fig:detector}), the convergence rate reduces to third order. It is worth noting that we found the convergence rate insensitive to the user-specified parameter $k_2$. The test with the M2-MP scheme yields seventh-order accuracy, with errors very close to those of the M1 scheme. For the solutions with discontinuities, the M0 scheme does not converge due to large dispersion errors, whereas first-order convergence with respect to the $L^1$ errors is observed with the M1 scheme. The M2-MP scheme also achieves nearly first-order convergence for the $L^1$ errors, whereas the M2-JS scheme converges the $L^1$ errors slightly slower. The $L^\infty$ errors for all schemes remain nearly constant with grid refinement. This is reasonable as the scalar gradients also scale with the grid spacing. The error convergence in the $L^\infty$ norm is generally considered unrealistic for solutions with discontinuities~\cite{Leveque1992}.

\subsection{Three-dimensional mixing layer} \label{3D_results}
In this section, to study the performance of the proposed schemes in under-resolved turbulent conditions, we evaluate them on a three-dimensional turbulent mixing-layer case with sharp gradients in density, temperature, and mass fractions. We assess the performance of these schemes regarding the mitigation of scalar excursions, numerical dissipation control, and the effect on scalar mixing. 

\subsubsection{Flow configuration} 
The flow configuration is adapted from \textcite{Matheou2016, Sharan2018}. It consists of a temporally evolving turbulent shear layer in a cubic domain with a side length of $L=4\pi$, spanning $[-2\pi,2\pi]\times[-2\pi,2\pi]\times[0,4\pi]$ in the $x$-, $y$-, and $z$-directions, respectively, and a triply periodic boundary condition. The flow was originally set up for incompressible flows with uniform density and a divergence-free constraint, whereas in the present study, we adapt it to a two-species compressible mixing layer with sharp gradients in density, temperature, and species mass fractions. The initial velocity fields in the streamwise ($x$), spanwise ($y$), and cross-stream ($z$) directions are, respectively,
\begin{align*}
    u &=      f(z) + a(z)\left[\sin{4x}+0.01\cos{20y}+r_u\right]\ , \\
    v &= \qquad\quad a(z)\left[\cos{4x}+0.01\sin{20y}+r_v\right]\ , \\
    w &= \qquad\quad a(z)\left[\sin{2x}+0.01\cos{40y}+r_w\right]\ ,
\end{align*}
where $f(z)$ is the initial mean velocity in the streamwise direction and varies from $-1$ to $1$ in the $z$-direction, given as 
\begin{equation}
    f(z) = \tanh{(40(\mathrm{modulo}(z,4\pi)-2\pi))} - \tanh{(40(\mathrm{modulo}(z,4\pi)-4\pi))}-\tanh{(40(\mathrm{modulo}(z,4\pi)))}.
    \nonumber
\end{equation}
The expressions in square brackets $[\cdots]$ represent initial perturbations imposed on the velocity field to induce shear-layer instabilities. It consists of streamwise and spanwise perturbations with forced wavenumbers and uniformly distributed perturbations, enforced by random numbers $r_u$, $r_v$, and $r_w$ in the range of $[-0.5,0.5]$ (see also Ref.~\cite{Sharan2019}). The perturbations are confined to the initial mixing layer via an amplitude function, 
\begin{equation}
    a(z) = 0.1 \exp{[-100(\mathrm{modulo}(z+\pi,2\pi)-\pi)^2]}\ .
    \nonumber
\end{equation}
With the development of shear-layer instabilities, the flow evolves into different structures, including initial Kelvin--Helmholtz vortices, uniformly mixed structures across the shear layer, and non-uniform three-dimensional large-scale structures. At these stages, the flow exhibits various turbulent mixing characteristics, providing a good testing bed to study the effect of numerical schemes on different mixing regimes. These features were previously observed for incompressible flows with uniform density~\cite{Sharan2018, Sharan2019}, and the current flow undergoes similar stages and results in similar flow characteristics, though with some discrepancies (e.g., asymmetry due to the non-uniform density). 

\begin{figure}[!htb]
    \centering
    \includegraphics[keepaspectratio=true]{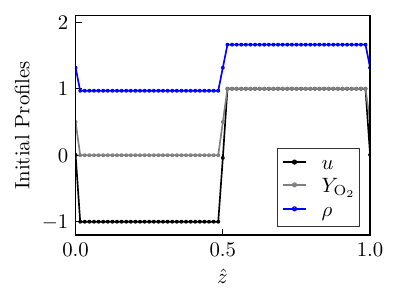}
    \caption{Initial profiles of the streamwise velocity, the $\mathrm{O_2}$ mass fraction, and the density along the cross-stream direction ($\hat{z}=z/L$).}
    \label{fig:initial} 
\end{figure}

The density and species mass fractions of $\mathrm{O_2}$ and $\mathrm{N_2}$ are initialised with the same profile as the mean streamwise velocity ($f(z)$), though scaled by respective maximum and minimum values,
\begin{align*}
    \rho &= 0.5(\rho_{\max} \quad\ + \rho_{\min \quad\ }) + 0.5(\rho_{\max} \quad\ - \rho_{\min} \quad\ )f(z)\ , \\
    Y_{\mathrm{O_2}} &= 0.5(Y_{\max,\mathrm{O_2}} + Y_{\min,\mathrm{O_2}}) + 0.5(Y_{\max,\mathrm{O_2}} - Y_{\min,\mathrm{O_2}})f(z)\ , \\
    Y_{\mathrm{N_2}} &= 1 - Y_{\mathrm{O_2}}\ ,
\end{align*}
where $\rho_{\max}=1.663$, $\rho_{\min}=0.969$, $Y_{\max,\mathrm{O_2}}=1$, and $Y_{\min,\mathrm{O_2}}=0$. No initial perturbations are added to the scalar fields. The initial profiles are shown in Figure~\ref{fig:initial}, illustrating the sharp jumps (one grid point) across the mixing layer, which are chosen to expose unboundedness issues and thus allow assessing the effectiveness of schemes to preserve boundedness. The initial pressure is uniform, $p=17.414$, and the mixture specific heat ratio is constant, $\gamma=\gamma_{\mathrm{O_2}}=\gamma_{\mathrm{N_2}}=1.4$, hence avoiding potential issues with spurious pressure oscillations. Accordingly, the initial Mach number is approximately $0.23$. Note that since the present study focuses on discretisation schemes for the convective terms, we chose the Euler equation as a model problem, which do not include a (physical) viscous term.

\subsubsection{Numerics}
The computational domain is discretised uniformly with $65^3$ grid points, resulting in significantly under-resolved simulations. We evaluate the numerical schemes with a dissipative numerical flux in Table~\ref{tab:testingmatrix}, i.e., M1, M2-JS, M2-MP, and M3-MP. Since the simulation with the non-dissipative scheme, M0, crashes at a very early stage ($\hat{t}\approx14.4$), the results are not discussed here. For the M2-JS, M2-MP, and M3-MP schemes, we apply the low-order dissipative flux to the continuity, energy, and species equations to address scalar gradients, but not to the momentum equation. This strategy is useful for reducing the effect of numerical dissipation on turbulence development and has been widely adopted in the literature, such as Refs.~\cite{Desjardins2008,Pettit2011PsiPhi,Subbareddy2017, Sharan2018}. The global bounds used with the M3-MP scheme are $[0,1]$ for the $\mathrm{O_2}$ mass fraction. It is worth noting that in M1, M2-JS, M2-MP, and M3-MP, the high-order artificial dissipation term acts only on the high-wavenumber range and can constitute a suitable sub-grid regularisation term for large-eddy simulations, as also demonstrated in Refs.~\cite{Larsson2007, Sciacovelli2021, Kord2023boundedness} for similar high-order dissipation terms. This interpretation of the dissipation term has been referred to in the literature as implicit large-eddy simulations \cite{Boris1992, Porter1994, Fureby1999, Ben2007}. The low-order dissipation terms, on the other hand, act on the entire wavenumber range but are activated only locally. The three-stage third-order TVD Runge--Kutta scheme~\cite{Shu1988} is employed with a Courant number of $0.01$.

\subsubsection{Flow characteristics} \label{subsubsec_validation}
\begin{figure}[!htb]
    \centering
    \hspace*{-1.0cm}
    \includegraphics[keepaspectratio=true,width=17cm]{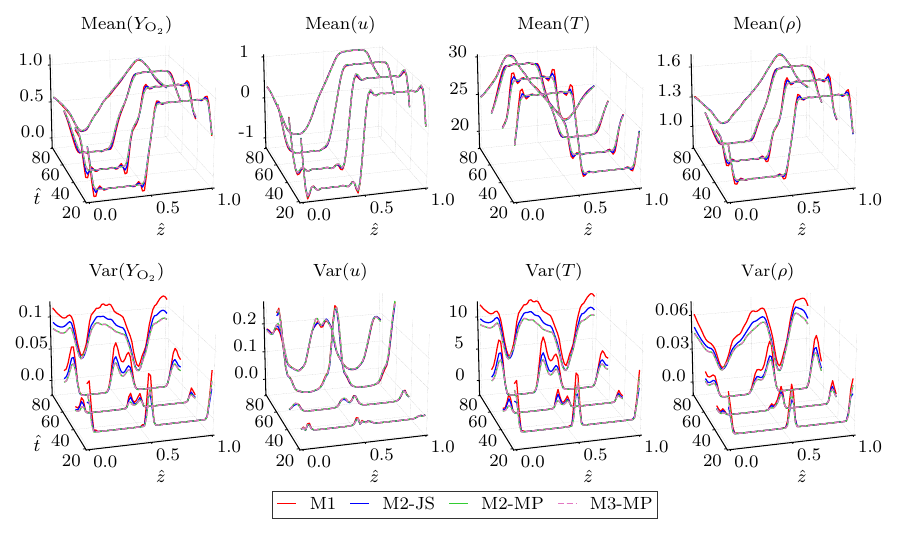}
    \caption{The means (averages on transverse planes, $\mathrm{Mean}(\cdot)$) and variances ($\mathrm{Var}(\cdot)$) of the $\mathrm{O_2}$ mass fraction, velocity, temperature, and density along the $z$-direction at time stages $\hat{t}=\numlist{20;40;60;80}$, from simulations using different numerical schemes. Note that the green solid line for M2-MP and the pink dashed line for M3-MP overlap.}
    \label{fig:validatequant}
\end{figure}

\begin{figure}[!htb]
    \centering
    \begin{subfigure}{\textwidth}
        \centering
        \hspace*{-1.0cm}\includegraphics[width=17cm]{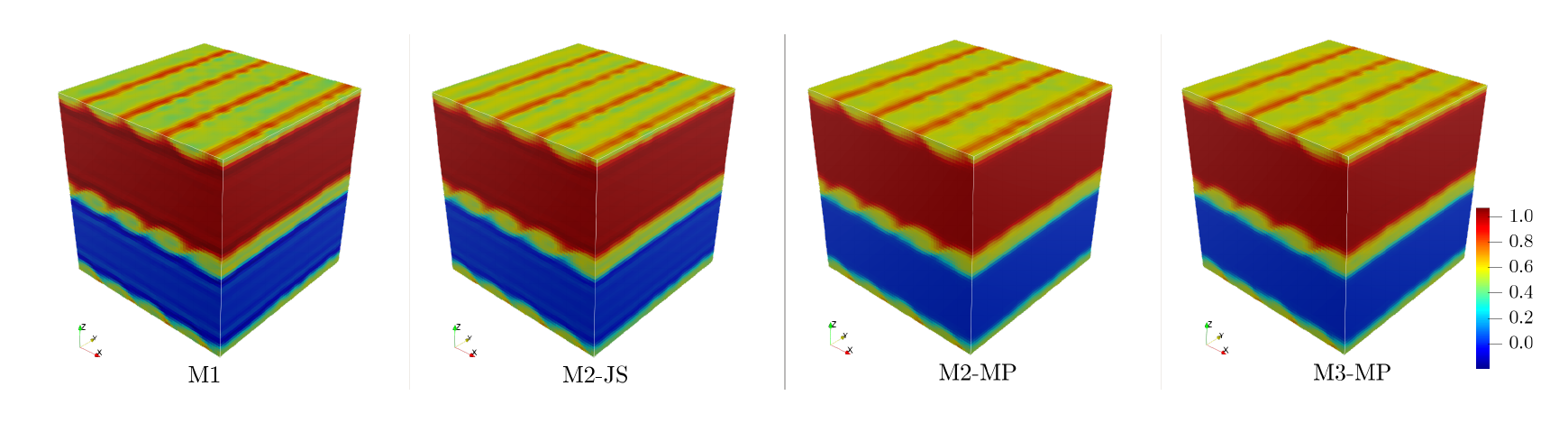}
        \caption{$\hat{t}=40$} \label{fig:overall40}
    \end{subfigure}
    
    \begin{subfigure}{\textwidth}
        \centering
        \hspace*{-1.0cm}\includegraphics[width=17cm]{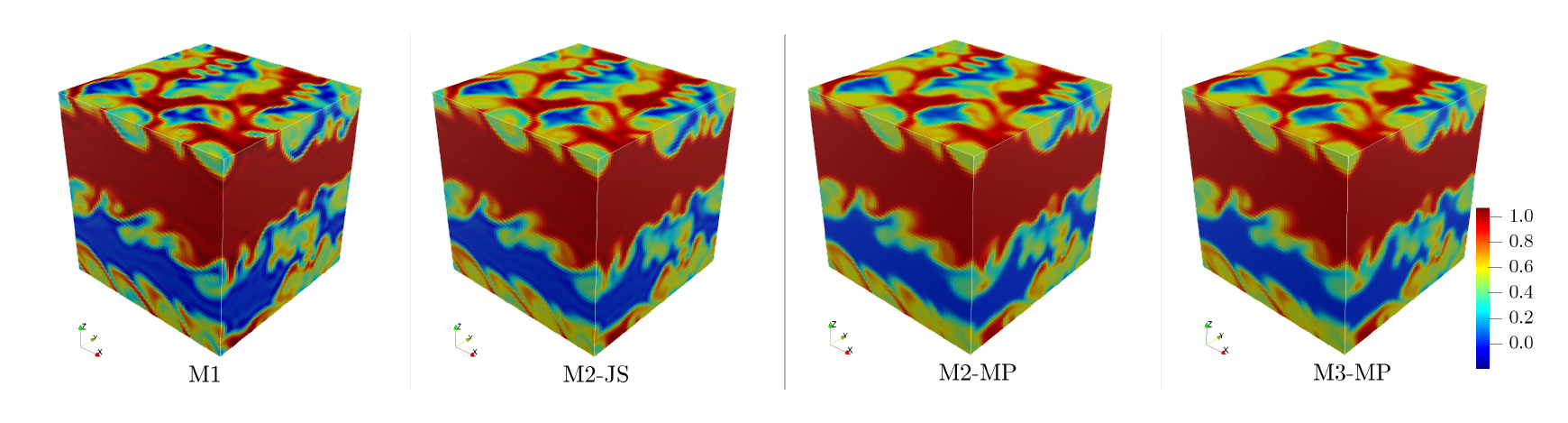}
        \caption{$\hat{t}=80$} \label{fig:overall80}
    \end{subfigure}
    
    \caption{Instantaneous visualisations of $\mathrm{O_2}$ mass fraction computed using different schemes at (a) $\hat{t}=40$ and (b) $\hat{t}=80$.}
    \label{fig:overall} 
\end{figure}

First, the flow evolution is briefly introduced. Figure~\ref{fig:validatequant} shows the means (averages on transverse planes) and the variances from the means of the $\mathrm{O_2}$ mass fraction, velocity, temperature, and density along the $z$-direction, i.e., across the mixing layer. The profiles at times $\hat{t}=t/\tau=\numlist{20;40;60;80}$ are plotted, where $\tau=1/4$ is a flow timescale determined by the initial convection velocity and the forced wavenumber~\cite{Sharan2018}. In addition, Figure~\ref{fig:overall} shows instantaneous visualisations of the $\mathrm{O_2}$ mass fraction distributions at $\hat{t}=40$ and $80$. 

As the flow evolves, Kelvin--Helmholtz instabilities emerge in the shear layer, and mixing happens between the two streams. At $\hat{t}=40$, the mixture within the mixing layer is largely uniform, as the $\mathrm{O_2}$ mass fraction shown in Figure~\ref{fig:overall40}. Subsequently, three-dimensional turbulent structures develop, and the mixture becomes stratified at $\hat{t}=80$ (Figure~\ref{fig:overall80}). The temporal evolution of the mean profiles for $Y_{\mathrm{O_2}}$, $T$, $\rho$, and $u$ (Figure~\ref{fig:validatequant}) consistently shows a gradual increase in the mixing and momentum layer thickness for all schemes.
However, only the M2-MP and M3-MP schemes successfully prevent the occurrence of under- and overshoots of scalars at the mixing-layer edges. This is also evident from the visualisations in Figure~\ref{fig:overall}, where these excursions manifest as stripes around the mixing layers for the M1 and M2-JS schemes. 

The temporal evolution of the variance profiles is also largely consistent between the different schemes, where the variance is highest for the M1 and M2-JS schemes and virtually the same for the M2-MP and M3-MP schemes. The scalar variance reaches a local minimum close to the second-stage mixing at $\hat{t}=40$, as discussed in Refs.~\cite{Sharan2018, Sharan2019}. This is due to the initial step profile and subsequently large variance in the mixing layer, which decays with the widening of the mixing layer (see also~Figure~\ref{fig:overall40}). The scalar variance increases again with the transition to turbulence at $\hat{t}=60$. The differences in magnitude can be directly attributed to the numerical diffusion induced by the numerical schemes as schemes including a low-order flux (M2-JS, M2-MP, and M3-MP) have significantly lower variance than the M1 scheme which includes only the high-order flux. The higher scalar variances in the mixing zones for the M2-JS scheme compared to the M2-MP and M3-MP schemes are likely caused by the unphysical oscillations and thus scalar under- and overshoots. The velocity variance profiles are virtually identical for all schemes, as the low-order fluxes are not applied to the momentum equation. 

Overall, all schemes predict a similar flow evolution, whereas evident discrepancies present in the scalar excursions and variances. Next, we will quantitatively evaluate the excursions and variances of the bounded schemes, M2-JS, M2-MP and M3-MP, with a comparison to the unbounded scheme, M1.

\subsubsection{Global scalar-excursion statistics} 
To evaluate the effectiveness of the bounded schemes in mitigating scalar oscillations, we compute the global scalar-excursion statistics, using the metrics proposed by \textcite{Matheou2016}. Here, we use the $\mathrm{O_2}$ mass fraction, $Y_{\mathrm{O_2}}$, and hence the lower and upper bounds are $0$ and $1$, respectively. We consider two metrics: the point-wise minimum and maximum $Y_{\mathrm{O_2}}$ in the computational domain, and the fraction of grid points that exceed the bounds. The first metric denotes maximum overshoots and maximum undershoots, akin to the maximum unboundedness error defined in Section~\ref{1D-boundedness}, whereas the second indicates the extent of unbounded regions.

\begin{figure}[!htb]
    \centering
    \includegraphics[keepaspectratio=true]{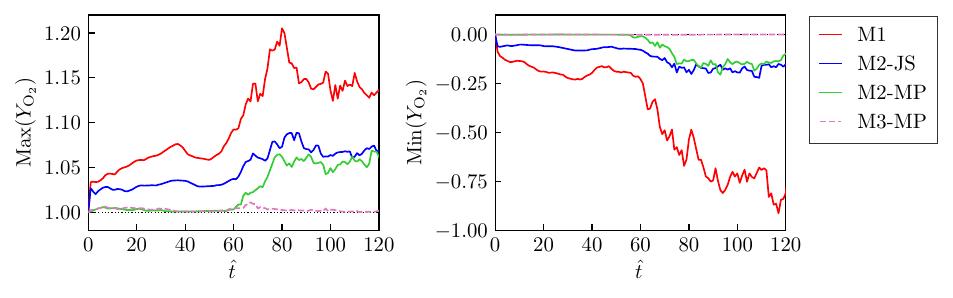}
    \caption{Temporal evolution of the global maximum over- and undershoots in the $\mathrm{O_2}$ mass fraction. Black dotted lines represent $Y_{\mathrm{O_2}}=1.0$ (Left) and $Y_{\mathrm{O_2}}=0.0$ (Right).}
    \label{fig:maxmin} 
\end{figure}

Figure~\ref{fig:maxmin} shows the histories of the maximum and minimum $Y_{\mathrm{O_2}}$ values. For the M1 and M2-JS, under-/overshoots develop right at the start of the simulation, and the further evolution shows a similar trend, though with a lager magnitude for the M1 scheme. In the early stage (up until $\hat{t}=60$), the M2-MP and M3-MP schemes lead only to very minor overshoots, which subsequently decay, and virtually no undershoots. As turbulence develops thereafter, under-/overshoots develop with the M2-MP scheme with magnitudes comparable to the M2-JS scheme. No undershoots and only marginal overshoots occur when the \textit{a posteriori} correction is employed (M3-MP). The results further indicate that the magnitude of the undershoots is significantly larger than for the overshoots, unlike in previous studies \cite{Matheou2016, Sharan2018} of incompressible flows in the same configuration. Overall, the results show that a scheme with only a high-order dissipation term (M1) is not effective in mitigating scalar excursions, contrary to schemes with a low-order flux term (M2-JS and M2-MP) which significantly reduce the such excursions. The M2-MP scheme is more effective in mitigating scalar excursions at earlier stages than M2-JS. The M3-MP scheme, featuring a low-order flux term and \textit{a posteriori} correction step, is most effective and prevents scalar excursions almost entirely.

\begin{figure}[tp]
    \centering
    \includegraphics[keepaspectratio=true]{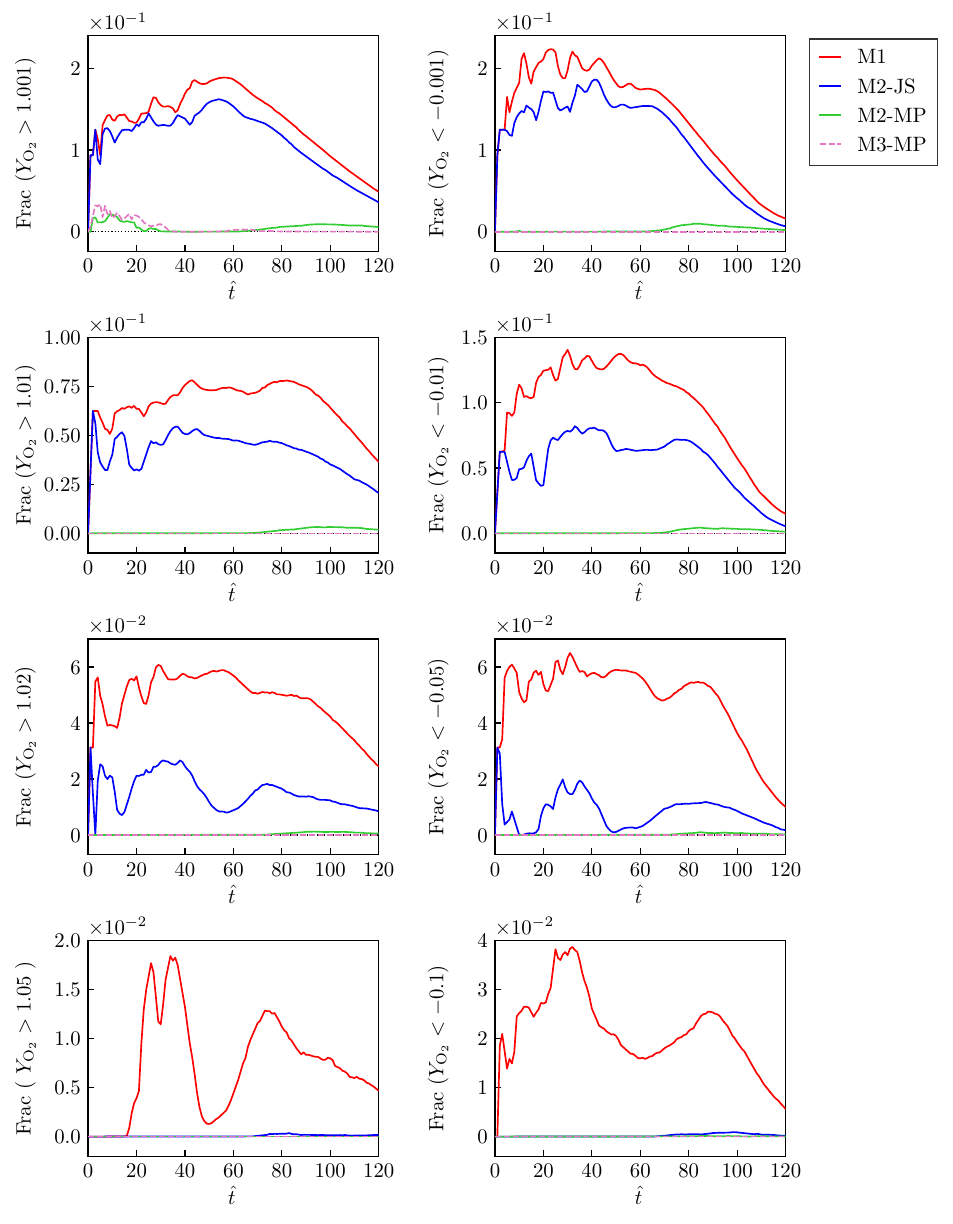}
    \vspace*{-8mm}
    \caption{Temporal evolution of the fractions of unbounded grid points that have excursions exceeding different threshold values. Black dotted lines represent $\mathrm{Frac}=0.0$.}
    \label{fig:vofex}
\end{figure}

Figure~\ref{fig:vofex} presents the temporal evolution of the fractions of grid points that have excursions. If the total number of such grid points is $N_b$, the fraction is computed as $N_b/N$, where $N$ is the total number of grid points in the domain. Different threshold values for the magnitude of excursions are considered, i.e., \numlist{1.001;1.01;1.02;1.05} for the overshoots, and \numlist{-0.001;-0.01;-0.05;-0.1} for the undershoots. The number of grid points with excursions for the M1 and M2-JS schemes is similar over the entire simulation -- the unbounded regions build up before $\hat{t}\approx60$ due to the large-scale entrainment of fluids as the mixing-layer development, and then diminish due to the increased numerical diffusion with the turbulent development. This observation is almost independent of the threshold value, particularly for smaller excursions; for example, both schemes yield a large fraction of grid points ($>5\%$) with minor excursions ($Y_{\mathrm{O_2}}>1.01$ and $Y_{\mathrm{O_2}}<-0.01$). However, the M2-JS scheme proves more effective in mitigating larger excursions. The M2-MP on the other hand leads to scalar excursions for a significantly smaller fraction of grid points ($<0.5\%$) of minor excursions, and only a marginal fraction of larger excursions. Adding the \textit{a posteriori} correction step in the M3-MP scheme, only a marginal fraction of grid points ($\approx 3\%$) show excursions larger than $0.1\%$ during the early flow development ($\hat{t}<30$) and no excursions after $\hat{t}\approx65$. The results show that a substantially improved boundedness is achieved with the M2-MP scheme relative to M2-JS, and a nearly strict boundedness with the \textit{a posteriori} flux correction (M3-MP).

Otherwise, we noted that, although not provided in the figure, applying the \textit{a posteriori} flux correction to the M1 and M2-JS schemes was not as effective as applying it to M2-MP. This is due to the considerably large fraction of unbounded points in simulations using M1 and M2-JS, and a single \textit{a posteriori} flux correction is not sufficient to mitigate the scalar excursions.

\subsubsection{Scalar and velocity spectra} 
\begin{figure}[tp]
    \centering
    \includegraphics[keepaspectratio=true]{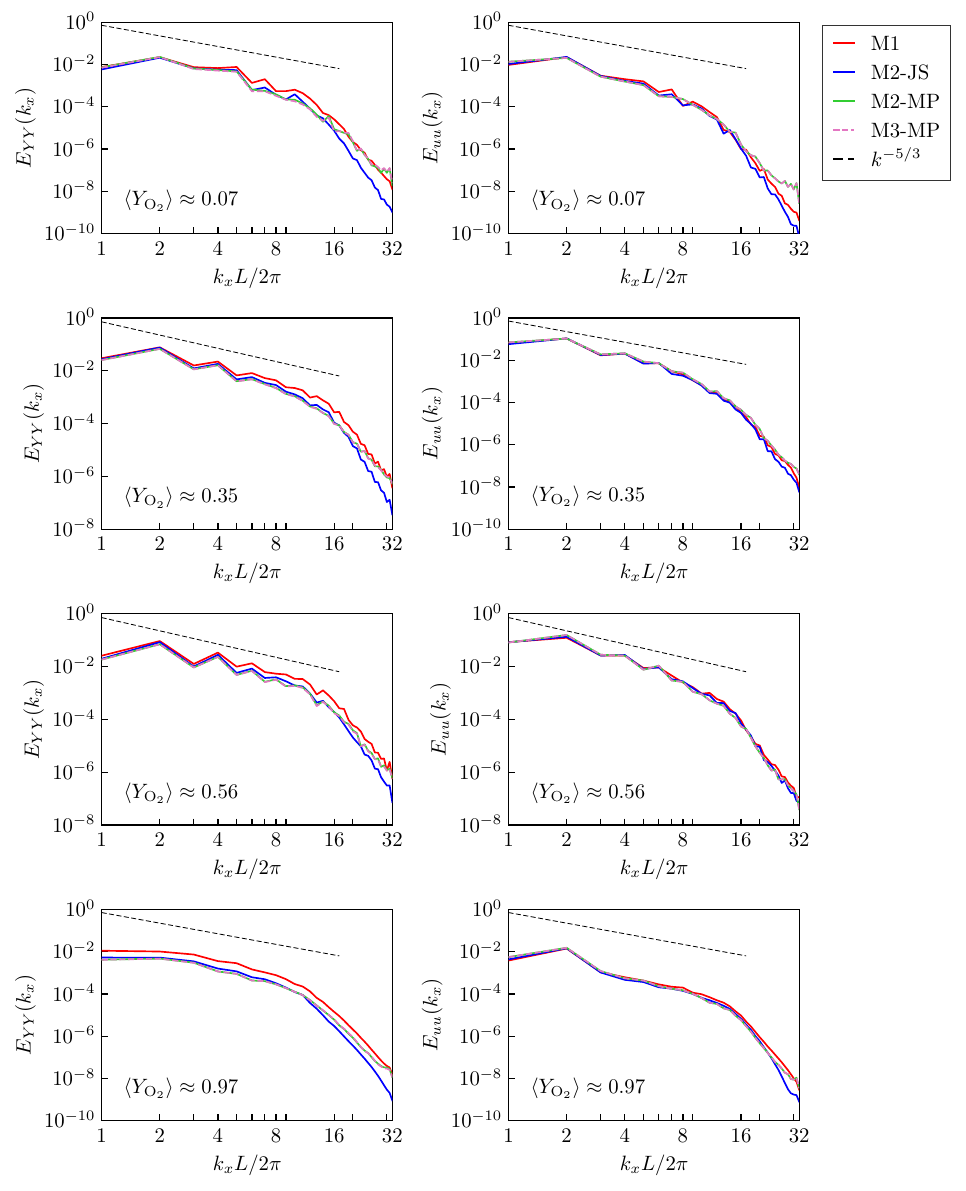}
    \vspace*{-8mm}
    \caption{One-dimensional mass fraction and velocity spectra at flow time $\hat{t}=80$ on different transverse planes.}
    \label{fig:spec}
\end{figure}

To evaluate the numerical dissipation introduced by the bounded components in M2-JS, M2-MP, and M3-MP, we analyse the scalar spectra next. Although the bounded flux, i.e., the low-order dissipative flux term, is not added to the momentum equations, the velocity spectra are also assessed. Figure~\ref{fig:spec} shows the one-dimensional scalar and velocity spectra along the streamwise ($x$) direction on several transverse planes at $\hat{t}=80$, which, on a given transverse plane $z=\hat{z}L$, are calculated as~\cite{Sharan2019}: 
\begin{align*}
    E_{YY}(k_x) &= \langle\ \lvert \mathrm{DFT}_{k_x}\{Y_{\mathrm{O_2}}(x,y)\}\rvert^2\ \rangle_y, \\
    E_{uu}(k_x) &= \langle\ \lvert \mathrm{DFT}_{k_x}\{u(x,y)\}\rvert^2\ \rangle_y,
\end{align*}
where $\mathrm{DFT}_{k_x}\{\cdot\}$ is the discrete Fourier transform in the $x$-direction and $\langle\cdot\rangle_y$ is the average over the $y$-direction. Different rows in Figure~\ref{fig:spec} correspond to different transverse planes, i.e., different $\hat{z}L$; however, instead of using the coordinates to denote the plane locations, we use the transverse average $\langle Y_{\mathrm{O_2}}\rangle$ to indicate the mixture state on the averaging plane. The selected locations, $\hat{z}=\numlist{0.25;0.39;0.5;0.72}$, are shown in Figure~\ref{fig:meanvar80} and correspond to $\langle Y_{\mathrm{O_2}}\rangle\approx\numlist{0.07;0.35;0.56;0.97}$, respectively. 

Both mass fraction and velocity spectra follow the $-5/3$ slope only for large scales within a narrow range of wavenumbers ($2<k_xL/2\pi<6$), but decay faster with an approximately $-3$ slope for $6<k_xL/2\pi<13$. At higher wavenumbers ($k_xL/2\pi>16$), the small-scale energies are dissipated due to regularisation by the high-order dissipation. The difference between the M1 scheme and the bounded schemes (M2-JS, M2-MP, and M3-MP) is attributed to dissipation from the low-order dissipative terms, which is the main metric we focus on next. For the scalar spectra, M2-JS introduces dissipation to a wide wavenumber range ($k_xL/2\pi>5$), and at $\langle Y_{\mathrm{O_2}}\rangle\approx0.97$ even the largest scales are affected. However, M2-MP introduces much less dissipation at high wavenumbers than M2-JS, although dissipation at larger scales is comparable between two. At the mixing-layer edges with near-bound mixtures ($\langle Y_{\mathrm{O_2}}\rangle\approx\numlist{0.07;0.97}$), the difference of scalar spectra between M2-MP and M1 is larger for $k_xL/2\pi\approx7$ than for other wavenumbers. On the middle planes with intermediate mixtures ($\langle Y_{\mathrm{O_2}}\rangle\approx\numlist{0.35;0.56}$), the larger difference shifts to higher wavenumbers ($k_xL/2\pi\approx16$). This occurs because near the mixing-layer edge, scalar oscillations tend to appear in large-scale structures, whereas in the mixing zones, fluctuations tend to emerge on smaller scales (cf.~Figure~\ref{fig:overall80}). This shift in the effective range of M2-MP's low-order dissipative term suggests that the M2-MP scheme adapts to different mixture compositions, which is not observed with M2-JS. Furthermore, the scalar spectra from M3-MP are indistinguishable from M2-MP, as the flux correction acts on minimal grid points (cf.~Figure~\ref{fig:vofex}). 

For the velocity spectra, the difference between the M1 scheme and the bounded schemes is significantly smaller than it is for the scalar spectra on all planes, which was anticipated since no changes were made to the momentum equations. The minor differences between the schemes are due to the fact that the velocity field is affected by density. At $\langle Y_{\mathrm{O_2}}\rangle\approx\numlist{0.07;0.97}$, M2-JS dissipates more energy at the highest wavenumbers than M2-MP. On the low-bound plane ($\langle Y_{\mathrm{O_2}}\rangle\approx0.07$), interestingly, M2-MP dissipates more slowly than M1. This is also seen in the scalar spectra and may result from the energy transported from larger scales or unphysical energy buildups. \textcite{Subbareddy2017} made a similar observation regarding the scalar spectra when evaluating bounded schemes (cf.~Fig.~3 in Ref.~\cite{Subbareddy2017}). On the planes with intermediate mixtures, the differences are relatively marginal.

In brief, numerical dissipation introduced to the scalar field by the low-order dissipative terms of the M2-JS and M2-MP schemes is not insignificant, whereas it has only marginal influence on the velocity field. However, the M2-MP scheme outperforms the M2-JS scheme, as M2-MP is less dissipative at small scales and can adaptively adjust the effective range of low-order dissipation. Lastly, the flux correction in the M3-MP scheme induces negligible dissipation.

\subsubsection{Scalar mixing}
Accurate predictions of the composition of the mixing field are vital in high-fidelity flow simulations, and in reacting simulations in particular. However, the prediction of turbulent mixing has been shown to be sensitive to numerical dissipation~\cite{Sharan2018}. In this section, we assess the impact of the additional numerical dissipation from the bounded components in M2-JS, M2-MP, and M3-MP on scalar mixing. 

\begin{figure}[!hbt]
    \centering
    \includegraphics[keepaspectratio=true]{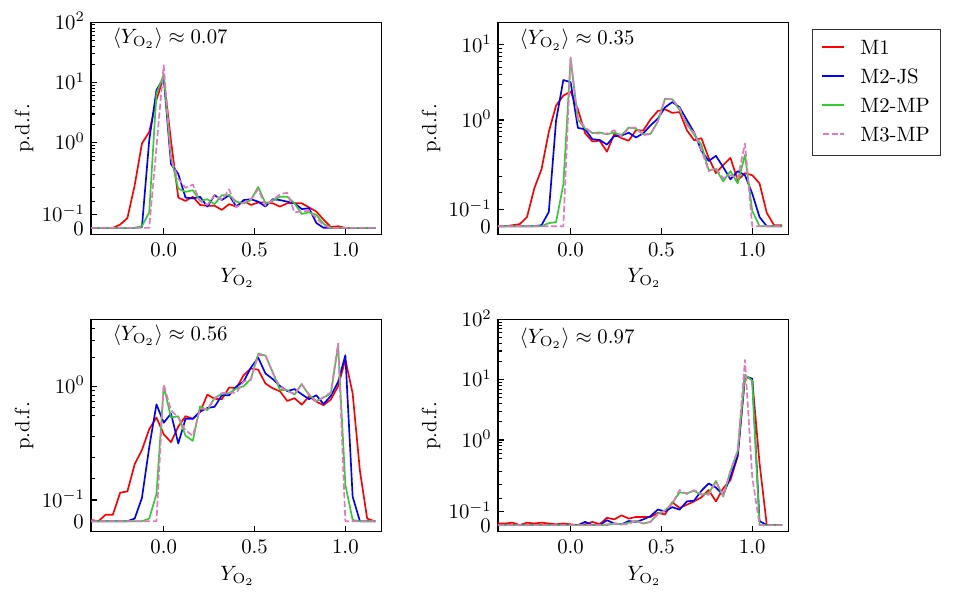}
    \vspace*{-8mm}
    \caption{The p.d.f.\ of $\mathrm{O_2}$ mass fraction on different transverse planes at flow time $\hat{t}=80$.}
    \label{fig:pdf80}
\end{figure}

\begin{figure}[!hbt]
    \centering
    \includegraphics[keepaspectratio=true]{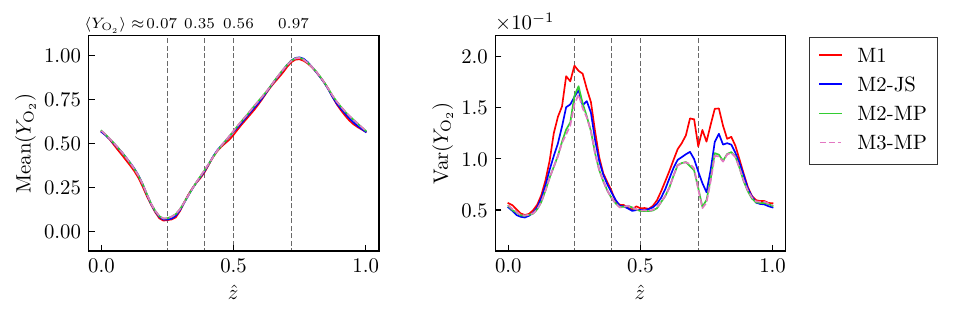}
    \vspace*{-8mm}
    \caption{The mean and mixed-fluid variance of $\mathrm{O_2}$ mass fraction along axis $\hat{z}=z/L$ at flow time $\hat{t}=80$.}
    \label{fig:meanvar80}
\end{figure}

Figure~\ref{fig:pdf80} shows the probability density function (p.d.f.) of $\mathrm{O_2}$ mass fraction on different transverse planes (marked in Figure~\ref{fig:meanvar80}) at $\hat{t}=80$. The vertical axes of Figure~\ref{fig:pdf80} are plotted on a logarithmic scale for values above $10^{-1}$, and otherwise, on the linear scale. Figure~\ref{fig:meanvar80} shows the mean and the mixed-fluid variance along axis $\hat{z}=z/L$ at $\hat{t}=80$. Note that the mixed-fluid variance is calculated as Refs.~\cite{Sharan2018, Sharan2019}, for a given $z$,
\begin{equation}
    \mathrm{Var}(Y_{\mathrm{O_2}})=\int_{\zeta}^{1-\zeta} (Y_{\mathrm{O_2}}-\langle Y_{\mathrm{O_2}}\rangle)^2\mathcal{P}(Y_{\mathrm{O_2}})\mathrm{d}(Y_{\mathrm{O_2}}),
    \nonumber
\end{equation}
where $\langle Y_{\mathrm{O_2}}\rangle$ and $\mathcal{P}(Y_{\mathrm{O_2}})$ respectively denote the mean and the p.d.f.\ of $\mathrm{O_2}$ mass fraction, and $\zeta=0.05$ is selected as a threshold between the mixed and unmixed fluids, and accordingly in this variance calculation merely the fluctuations in the mixed fluids are accounted for and the impact of unbounded values is excluded. The mean indicates a local average of the mixture state, determined by large-scale structures, while the mixed-fluid variance represents local scalar fluctuations, driven by small-scale processes~\cite{Sharan2019}.

We focus on the impact of the additional numerical dissipation introduced by the bounded terms in M2-JS, M2-MP, and M3-MP, which is shown by the difference between M1 and these bounded schemes. At the mixing-layer edges ($\langle Y_{\mathrm{O_2}}\rangle\approx\numlist{0.07;0.97}$), the fluid elements exceeding the $Y_{\mathrm{O_2}}$ bounds are reasonably smeared out by the additional diffusion of the bounded schemes. However, these bounded schemes also slightly affect the mixed fluids in the remaining range of $Y_{\mathrm{O_2}}$ (around $Y_{\mathrm{O_2}}>0.2$ on plane $\langle Y_{\mathrm{O_2}}\rangle\approx0.07$ and $Y_{\mathrm{O_2}}<0.8$ on plane $\langle Y_{\mathrm{O_2}}\rangle\approx0.97$). This is indicated by the more contracted p.d.f.\  with the bounded schemes compared to M1. Despite the low probability of fluids in these ranges, this effect leads to evident reductions in the mixed-fluid scalar variance, particularly on plane $\langle Y_{\mathrm{O_2}}\rangle\approx0.97$, as illustrated in Figure~\ref{fig:meanvar80}. The larger variance reduction on (and around) this plane is likely due to the higher numerical dissipation introduced by all the bounded schemes, as seen in Figure~\ref{fig:spec}. Nevertheless, given that M1 exhibits large scalar excursions at these mixing-layer edges, unphysical scalar oscillations potentially exist within the global bounds, and the variance reduction by the bounded schemes might be a reasonable result of removing such errors. This reasoning may also explain why M2-JS, despite its higher numerical dissipation (Figure~\ref{fig:spec}), still results in slightly higher variance compared to M2-MP, as, similar to M1, M2-JS may also produce unphysical oscillations within the global bounds. On the planes in the mixing zones ($\langle Y_{\mathrm{O_2}}\rangle\approx\numlist{0.35;0.56}$), M1 shows a non-uniform composition with a peak at around $0.5$. With the bounded terms, the fluid elements exceeding the mass fraction bounds are removed, except for M2-JS, where excursions remain. All bounded schemes lead to slight contractions of the p.d.f.\ distributions in the mixed-fluid range and small shifts of the p.d.f.\ peak on plane $\langle Y_{\mathrm{O_2}}\rangle\approx{0.56}$. Despite that, these changes minimally affect the mixed-fluid variance, as shown in Figure~\ref{fig:meanvar80}. Otherwise, the mean profiles in Figure~\ref{fig:meanvar80} from all schemes nearly overlap, indicating that the overall mixing state is insensitive to the low-order dissipation.

\begin{figure}[!hbt]
    \centering
    \includegraphics[keepaspectratio=true]{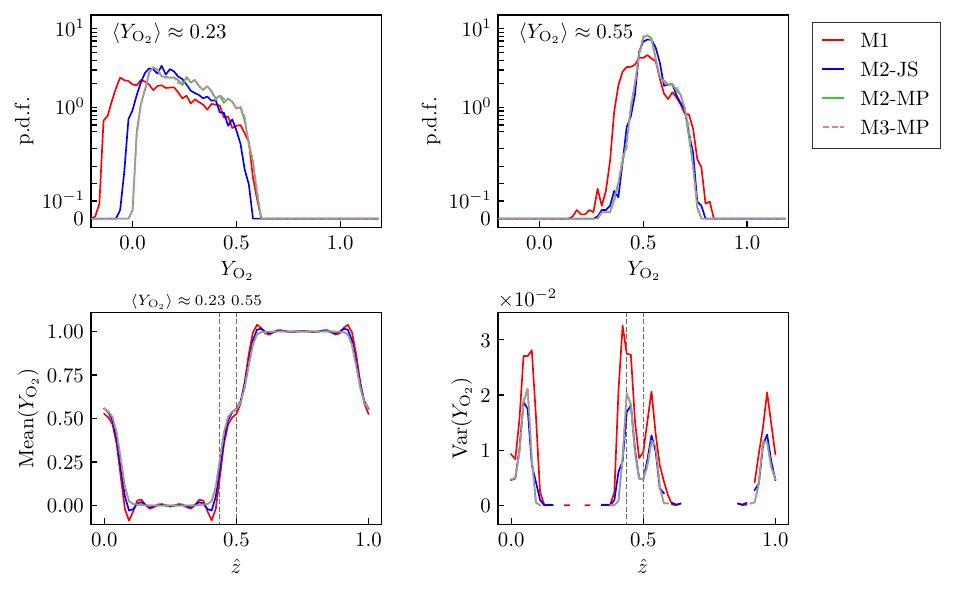}
    \vspace*{-8mm}
    \caption{The p.d.f.\ of $\mathrm{O_2}$ mass fraction on different transverse planes at flow time $\hat{t}=48$.}
    \label{fig:pdfmeanvar48}
\end{figure}

\textcite{Sharan2018} pointed out that the mid-plane scalar p.d.f.\ and mixed-fluid scalar variance are more sensitive to numerical diffusion at the development stage of the shear flow when the mixture composition is near-uniform, compared to the fully turbulent stage described above. Therefore, to assess the impact of the additional numerical dissipation on a different mixing regime, we next analyse the scalar mixing statistics at $\hat{t}=48$, where the mid-plane mixture shows a more uniform composition. Figure~\ref{fig:pdfmeanvar48} (top row) shows the $Y_{\mathrm{O_2}}$ p.d.f.\ on the planes with $\langle Y_{\mathrm{O_2}}\rangle\approx\numlist{0.23;0.55}$, which are close to the mixing-layer edge and on the mid-plane, respectively. The locations of the planes are indicated in the bottom-row plots, showing the mean and variance along the $z$-direction. The mixed-fluid variance is calculated similarly to that in Figure~\ref{fig:meanvar80}, yet with a relaxed threshold $\zeta=0.01$ as the extent of mixing is limited at $\hat{t}=48$. The gaps between the line segments in the variance plot are due to the absence of mixed fluids ($Y_{\mathrm{O_2}}<0.01$ or $Y_{\mathrm{O_2}}>0.99$).

At $\hat{t}=48$, the mixing zone is primarily characterised by two-dimensional vortices spanning in the $y$-direction, similar to the structures at $\hat{t}=40$ visualised in Figure~\ref{fig:overall40}. Therefore, the p.d.f.\ on the mid-plane shows a unimodal distribution around $Y_{\mathrm{O_2}}=0.5$ consistently for all schemes. Comparing the bounded schemes to the M1 scheme in Figure~\ref{fig:pdfmeanvar48}, it is evident that the additional dissipation from the low-order terms induces enhanced mixing to the mid-plane fluids, such that the mixture compositions become more uniform and the scalar variances are reduced in all the bounded schemes. On the plane close to the mixing-layer edge, the low-order dissipation terms are largely activated to suppress spurious oscillations and excursions, thus leading to significant variance reductions. Therefore, both the artificially enhanced mixing and removed oscillations cause the scalar variance reductions. The clear difference between the bounded schemes and M1 in both p.d.f.\ and variance indicates that the impact of numerical dissipation is larger at this stage than at $\hat{t}=80$. 

Although the impact on mixing due to the bounded terms may be non-negligible, at earlier stages in particular, it is promising that the M2-MP scheme produces comparable influences to the M2-JS scheme but brings significant improvements in scalar boundedness. Considering that the effect on mixing becomes gradually small at later stages, it is uncertain how significantly the accuracy of a reacting-flow simulation can be affected, necessitating further assessments, which are currently ongoing.

\section{Conclusion} \label{Sec_con}
In this work, we developed formulations for high-order finite-difference methods to preserve scalar boundedness without predefined bounds while maintaining high accuracy and low numerical dissipation. We were specifically concerned with the scalar transport equations for species mass fractions in turbulent compressible multi-component flow simulations. The proposed schemes were examined through numerical experiments on 1D scalar wave advection and 3D turbulent mixing-layer problems involving sharp scalar gradients and under-resolved conditions, focusing on scalar boundedness, numerical diffusivity, and the impact on turbulent mixing.

The proposed formulations were built upon a numerical flux framework that augments a non-dissipative numerical flux with an explicit dissipative numerical flux. The non-dissipative term uses a split numerical flux of a high-order central-difference scheme that we identified in a previous study as stable and physically consistent. The dissipative term adaptively switches between a high-order local Lax-Friedrichs flux and a low-order bounded flux. The high-order dissipation eliminates high-wavenumber modes in under-resolved simulations, the low-order dissipation ensures scalar boundedness, and the switch localises the low-order dissipation near sharp scalar gradients. We constructed two dissipative fluxes: one follows Jameson's artificial viscosity method, using a density-based sensor (M2-JS), and the other employs a monotonicity-preserving limiter to switch between high- and low-order dissipative fluxes (M2-MP). Neither scheme requires inputs on scalar bounds, allowing to preserve scalar bounds determined by local conditions rather than prescribed global physical bounds such as $[0,1]$ for species mass fractions. Last, we applied an \textit{a posteriori} flux correction step to ensure strict scalar boundedness. 

The numerical experiments demonstrated good numerical stability for both M2-JS and M2-MP schemes, and showed that M2-MP preserved high-order accuracy, while M2-JS caused a degradation in the order of accuracy. The species mass fractions were almost bounded with M2-MP, exhibiting very limited excursions and practically no oscillations in density and temperature fields, which was substantially improved compared to M2-JS. Additionally, M2-MP introduced less numerical dissipation than M2-JS. In the turbulent case, the low-order numerical dissipation did not affect the velocity fields in either scheme as the low-order fluxes were not applied to the momentum equations, whereas the scalar mixing fields were clearly affected for both schemes, with scalar fluctuations smeared out near the mixing-layer edges. This is reasonable and probably inevitable for the present highly under-resolved case, given the otherwise high excursions. Nevertheless, M2-MP is promising because both 1D and 3D tests demonstrated that it outperformed M2-JS, which is based on the well-established Jameson’s artificial viscosity method. By preserving accuracy, offering better scalar boundedness, being more solution-adaptive, and being less dissipative at small scales, M2-MP demonstrated superior performance, and is therefore the preferred scheme. Furthermore, applying the \textit{a posteriori} flux correction to M2-MP achieved nearly strict scalar boundedness within the prescribed global bounds with almost no influence on the flow fields, and thus this correction is recommended as a fail-safe option.

\section*{Acknowledgement}
This work was supported by the Australian Government through the Australian Research Council's Discovery Projects funding scheme (project DP200103535). A pool of computational resources was provided by the Australian Government through the Pawsey Supercomputing Centre and the National Computational Infrastructure under the National Computational Merit Allocation Scheme, and by the University of New South Wales.

\printbibliography

\end{document}